\documentclass[12pt]{article}

\usepackage{amsmath}
\usepackage{amsfonts,dsfont,mathrsfs}
\usepackage{amssymb}
\usepackage{graphics,graphicx,tikz,tensor}
\usetikzlibrary{arrows,decorations.pathmorphing,patterns}
\usepackage[linktocpage]{hyperref}
\hypersetup{
	colorlinks=true,
	linkcolor=blue,
	filecolor=magenta,      
	urlcolor=blue,
	citecolor=magenta
}
\numberwithin{equation}{section} 
\usepackage{caption,subcaption}
\usepackage{cancel}
\usepackage{tabu}
\usepackage{empheq}

\usepackage[noadjust]{cite}

\newcommand{\ft}[2]{{\textstyle\frac{#1}{#2}}}
\newcommand{\nn}{\nonumber}

\def\be{\begin{equation}}
	\def\ee{\end{equation}}
\def\bea{\begin{align}}
	\def\eea{\end{align}}
\def\beaq{\begin{eqnarray}}
	\def\eeaq{\end{eqnarray}}

 \def\D{\Delta}

\usepackage{color}

\begin{document}

\begin{titlepage}
	%\hfill \hbox{????}
	%\vskip 0.1cm
	%\hfill  \hbox{????}
	%\vskip 1.5cm
	\begin{flushright}
	\end{flushright}
	\vskip 1.0cm
	\begin{center}
		{\Large \bf ``Waveforms'' at the Horizon}
		\vskip 1.0cm {A. Cipriani,$^a$ F. Fucito$^a$, C. Heissenberg,$^b$ J.F.  Morales,$^a$ R. Russo$^c$} \\[0.7cm]
		{\it \small$^a$Dipartimento di Fisica, Universit\`a di Roma ``Tor Vergata'' 
			\& 
			\\ Sezione INFN ``Roma Tor Vergata'',
			Via della Ricerca Scientifica 1, 00133, Roma, Italy}\\[0.5cm]
		{\it \small$^b$Institut de Physique Th\'eorique, CEA Saclay, CNRS, Universit\'e Paris-Saclay, \\
			F-91191, Gif-sur-Yvette Cedex, France}\\[0.5cm]
		{\it \small$^c$School of Mathematical Sciences,
			Queen Mary University of London, \\
			Mile End Road, London, E1 4NS, United Kingdom.}
	\end{center}
	
	\vspace{5pt}
	
	\begin{abstract}
		We study perturbations induced by a light particle scattering off a Schwarzschild black hole. Exploiting recent results for the wave propagation in this geometry, we derive the fields that this process induces on the horizon to leading order in the post-Minkowskian (PM) regime, when the light probe is far from the black hole. We then use these results to calculate the fluxes of energy and angular momentum that enter the black hole. We consider the effects due to gravitational, electromagnetic and scalar radiation, finding agreement with recent computations of the absorbed energy, while the absorbed angular momentum provides a new PM result.
	\end{abstract}
\end{titlepage}

\tableofcontents

\section{Introduction}

The flourishing of gravitational-wave astronomy during the last decade \cite{LIGOScientific:2016aoc,LISA:2017pwj,LIGOScientific:2017vwq,KAGRA:2021vkt,LISA:2022kgy,LIGOScientific:2025slb}  has put the gravitational two-body problem in the spotlight, stimulating renewed interest in approximation methods that can be used to gain analytical control on its dynamics. Among the various approaches,
two domains that have been recently witnessing  interesting developments are the post-Minkowskian (PM) approximation and black hole perturbation theory. The first strategy applies in the weak-field regime, when the two interacting objects are far apart, and retains an exact dependence on the objects' velocities and masses. It applies in a particularly neat way to  gravitational scatterings, in which the asymptotic states are initially freely moving on a Minkowski background, whose dynamics can be naturally expressed in an effective field theory (EFT) approach in terms of scattering amplitudes \cite{Bjerrum-Bohr:2018xdl,Cheung:2018wkq,Kosower:2018adc,Bern:2019nnu,KoemansCollado:2019ggb,Bern:2019crd,Bjerrum-Bohr:2019kec,Bern:2020gjj,Cristofoli:2020uzm,Parra-Martinez:2020dzs,DiVecchia:2021ndb,Bern:2021dqo,Herrmann:2021lqe,DiVecchia:2021bdo,Herrmann:2021tct,Bjerrum-Bohr:2021din,Cristofoli:2021vyo,Brandhuber:2021eyq,Bern:2021yeh,DiVecchia:2023frv,Brandhuber:2023hhy,Herderschee:2023fxh,Elkhidir:2023dco,Georgoudis:2023lgf,DeAngelis:2023lvf,Brandhuber:2023hhl,Georgoudis:2023eke,Georgoudis:2023ozp,Georgoudis:2024pdz,Brunello:2024ibk,Alessio:2024onn,Brunello:2025eso,DeAngelis:2025vlf,Ivanov:2025ozg} and perturbative worldline calculations \cite{Kalin:2020mvi,Kalin:2020fhe,Mogull:2020sak,Dlapa:2021npj,Jakobsen:2021lvp,Dlapa:2021vgp,Kalin:2022hph,Jakobsen:2022psy,Dlapa:2022lmu,Driesse:2024xad,Bern:2024adl,Bern:2025wyd,Driesse:2026qiz}. The second strategy assumes instead that one of the two objects is much heavier, so fluctuations of spacetime generated by  the motion of the  light particle can be viewed as a perturbation of the background geometry.    
 This strategy has been applied to the study of compact binary systems in the Post Newtonian (PN) approximation via the MST method (see \cite{Mino:1997bx,Blanchet:2013haa} for reviews and references therein) and more recently via gauge-theory-inspired techniques 
\cite{Aminov:2020yma,Bianchi:2021xpr,Bianchi:2021mft,Bonelli:2021uvf,Bonelli:2022ten,Consoli:2022eey,Bautista:2023sdf,Aminov:2023jve,Fioravanti:2025bts}.  Quasi-circular binaries were studied in \cite{Fucito:2023afe}, and unbounded scattering processes in \cite{Fucito:2024wlg}, using black hole perturbation theory, and compared against  \cite{Bini:2018ylh,Bini:2020zqy} obtained via the multipolar post-Minkowskian method.

From a theoretical standpoint, an important feature of these complementary methods is that they can provide easier access points to different physical phenomena that characterize the gravitational dynamics. PM methods, which rely on an expansion in the gravitational coupling, are well suited to calculate the deflection angle \cite{Bjerrum-Bohr:2018xdl,Cheung:2018wkq,Bern:2019nnu,KoemansCollado:2019ggb,Bern:2019crd,Bjerrum-Bohr:2019kec,Bern:2020gjj,Cristofoli:2020uzm,Parra-Martinez:2020dzs,Kalin:2020mvi,Kalin:2020fhe,Damour:2020tta,DiVecchia:2021ndb,Bern:2021dqo,DiVecchia:2021bdo,Herrmann:2021tct,Bjerrum-Bohr:2021din,Brandhuber:2021eyq,Dlapa:2021npj,Bern:2021yeh,Dlapa:2021vgp,Dlapa:2022lmu,Driesse:2024xad,Bern:2024adl,Bern:2025wyd,Driesse:2026qiz}, including also dissipative effects such as radiation-reaction, and  the scattering waveform at infinity for generic velocities \cite{Kovacs:1977uw,Kovacs:1978eu,Jakobsen:2021smu,Mougiakakos:2021ckm,Brandhuber:2023hhy,Herderschee:2023fxh,Elkhidir:2023dco,Georgoudis:2023lgf,Caron-Huot:2023vxl,Georgoudis:2023eke,Georgoudis:2023ozp,Georgoudis:2025vkk,Brunello:2025eso}. Conversely, working perturbatively in the mass ratio allows for more control on non-analytic terms in the coupling, such as the tails and tails of tails logarithmic terms recently characterized in~\cite{Fucito:2024wlg} (see also \cite{Alessio:2024onn,Ivanov:2025ozg,Bini:2025ltr,Bini:2025bll}) and on related contributions that show up as infrared divergences on the EFT side.
A  PM formula that, at each order in the gravitational coupling, is exact in the velocity was recently derived in \cite{Cipriani:2025ikx}  in the probe approximation. Of course the two approaches can be compared in the common regime of validity, as, for instance, done in~\cite{Fucito:2024wlg} for the scattering waveform in the regime where both the deflection angle and the mass ratio are small. Then the amplitudes perspective can then be used to understand how to move away from the latter approximation \cite{Damour:2019lcq,Heissenberg:2025fcr}, while the self-force approach \cite{Poisson:2011nh,Pound:2021qin} can be used to go beyond the large impact parameter regime \cite{Adamo:2021rfq,Adamo:2022rmp,Adamo:2022qci,Adamo:2023cfp}. For instance, using this approach, \cite{Bini:2026cpw} derived the PN expanded radiated energy from a radially infalling particle in a Schwarzschild spacetime.

Moreover, by its very nature, the PM EFT does not capture the ``microscopic'' properties of the objects entering the two-body process, which appear as unspecified Wilson coefficients in the effective Lagrangian and need to be fixed by matching to a more fine-grained description. This is of course provided by black hole perturbation theory. In~\cite{Goldberger:2020wbx,Jones:2023ugm}, the two approaches were combined to calculate dissipative effects in a two-body scattering due to horizon absorption, leading to new PM predictions for the amount of energy that is dissipated due to this effect in a scattering of Schwarzschild black holes. See also \cite{Bautista:2024emt} for a generalization of this result to Kerr black holes and~\cite{Gatica:2025uhx} for an extension of the EFT approach to  generic, compact, spinning bodies.

In this work, we study the scattering of a light particle off a spinless black hole focusing on trajectories with large impact parameter. We use black hole perturbation theory and compute the energy and angular momentum absorbed by the black hole, by looking at the field sourced by the light particle moving in the Schwarzschild geometry. We exploit the recent results of~\cite{Cipriani:2025ikx} for the wave propagation to explicitly calculate the field perturbation that such a motion induces \emph{on} the black hole horizon, which we dub the ``waveform'' at the horizon. We apply the standard Teukolsky formalism~\cite{Teukolsky:1972my} which allows us to consider gravitational, electromagnetic and scalar perturbations in a unified way. Exactly as it happens for the usual waveform, the results for the perturbation at the horizon are naturally written in terms of the the spin-weighted spherical harmonics. However there exists a key difference between the two cases: for the waveform at infinity all harmonics contribute to leading order in the PM expansion and only in the PN limit the low harmonics become dominant, instead, for the waveforms at the horizon, each harmonic is weighted by a factor of $G^{\ell+1}$. Thus the lowest-$\ell$ contribution captures the leading PM result for arbitrary velocities.

We  then substitute the waveform at the horizon in the relevant Noether currents to calculate the energy and angular momentum fluxes absorbed by the black hole. We obtain explicit results to leading order for large impact parameter, i.e.~at the first nontrivial PM order. In the case of the absorbed energy, we provide new expressions for the spectral absorption rate and recover the results of~\cite{Goldberger:2020wbx,Jones:2023ugm} for the total increase in the black hole mass. This shows explicitly that, as expected, a balance law holds: the gravitational energy flowing into the horizon is equal to the change in the mass of the black hole. It is natural to expect that the same is true for our new results about the absorption of angular momentum and that the final state after the scattering involves Kerr, rather than a Schwarzschild, black hole with spin given at leading order in $G$ by~\eqref{eq:Jabs}. Indeed, this is supported by a check with the results of \cite{Goldberger:2020wbx} in the nonrelativistic limit. As is clear from the comparison with the EFT approach, the leading PM formulae obtained in this paper are exact in the masses, even if they are obtained in the probe limit, and one can obtain the leading PM energy and angular momentum absorbed by the light particle simply by swapping the role of the objects in the final results.

The paper is organized as follows. In Section~\ref{sec:wave_equations_all}, we review the equations and solutions governing the radial propagation of spin-$s$ waves in the Schwarzschild geometry. In Section~\ref{sec:computation_all} we compute the absorbed energy and angular momentum at leading PM order. Section~\ref{sec:conclusions} contains our conclusions. Three appendices contain some technical details. 

\section{Spin-$s$ waves on Schwarzschild geometry} \label{sec:wave_equations_all}

In this section, we review the wave equations describing scalar, electromagnetic and linearized gravitational perturbations propagating on a Schwarzschild geometry and discuss their solutions. We then study the waves emitted by the motion of point particles along the equatorial plane and display formulae for the energy and angular momentum emitted towards infinity and into the black hole horizon.

\subsection{Equations of motion} \label{sec:eoms}

We consider the Schwarzschild metric
\begin{equation}\label{eq:Schwarzschild}
	ds^2 =  - f(r) \,dt^2 + \frac{dr^2}{f(r)} + r^2 d\theta^2 + r^2 (\sin\theta)^2 d\phi^2\,,
\end{equation}
with
\begin{equation}
	f(r) = 1 - \frac{2 M}{r}\,,\qquad
	M = G M_\text{BH}\,,
\end{equation}
where $M_\text{BH}$ is the mass  of the black hole.
We are interested in a perturbation of the metric \eqref{eq:Schwarzschild},\footnote{A $\delta$ before a generic quantity refers to its expression at the linear order in the perturbation $\delta g_{\alpha \beta}$.} 

\begin{equation}
	\mathsf{g}_{\alpha\beta} = g_{\alpha\beta} +\delta g_{\alpha\beta}\,,\qquad \delta g_{\alpha\beta}  = h_{\alpha\beta}\,, \end{equation}
generated by the motion of a light particle of mass $\mu$. Here $\mathsf{g}_{\alpha\beta}$ is the full metric, $g_{\alpha\beta}$ is the background Schwarzschild metric given by \eqref{eq:Schwarzschild} and $h_{\alpha\beta}$ is the perturbation. Similarly, for the Riemann and Weyl tensors one has
\begin{equation}
	\mathsf{R}_{\alpha\beta\gamma\delta} = R_{\alpha\beta\gamma\delta} + \delta R_{\alpha\beta\gamma\delta}+\cdots\,,
	\qquad
	\mathsf{C}_{\alpha\beta\gamma\delta} = C_{\alpha\beta\gamma\delta} + \delta C_{\alpha\beta\gamma\delta}+\cdots
\end{equation}
up to linear order in the perturbation, or in the mass $\mu$ of the probe.
The Penrose scalars are given by
\begin{equation}
	\mathsf{\Psi}_0 = \delta C_{\mu\nu\rho\sigma} \ell^\mu m^\nu \ell^\rho m^\sigma + \cdots\,,
	\qquad
	\mathsf{\Psi}_4 = \delta C_{\mu\nu\rho\sigma} n^\mu \bar{m}^\nu n^\rho \bar{m}^\sigma + \cdots\,,
\end{equation}
where the \emph{background} tetrad vectors are %\eqref{eq:Kinnersley}
\begin{equation}\label{eq:Kinnersley}
	\ell = \frac{1}{f(r)}\,\partial_t + \partial_r\,,\qquad
	n = \frac{1}{2}\left(
	\partial_t  - f(r) \partial_r
	\right),\qquad
	m = \frac{1}{\sqrt{2}\,r}\left(
	\partial_\theta + \frac{i}{\sin\theta}\,\partial_\phi
	\right)
\end{equation}
and $\bar{m}$ is the complex conjugate of $m$. In Appendix~\ref{sec:app_A} we report tetrad conventions and other details on the background geometry.
By expanding in a similar way the Einstein tensor
\begin{equation}
	\mathsf{G}_{\alpha\beta} = \mathsf{R}_{\alpha\beta} - \frac{1}{2}\,\mathsf{g}_{\alpha\beta}\mathsf{R} = G_{\alpha\beta} + \delta G_{\alpha\beta} + \cdots \,,
\end{equation}
we obtain the linearized Einstein equations\footnote{More precisely, the Einstein equations read
$
	\mathsf{G}_{\alpha\beta} = 8\pi G\, \mathsf{T}_{\alpha\beta}
$
and, when the only source is the light particle,
\begin{equation*}
	\mathsf{T}^{\alpha\beta}(x) =\frac{\mu}{\sqrt{-\mathsf{g}(x)}} \int  \frac{\delta^{(4)}(x-x(\lambda)) \partial_\lambda x^\alpha(\lambda) \partial_\lambda x^\beta(\lambda) }{\sqrt{-\partial_\lambda x^\rho(\lambda)\mathsf{g}_{\rho\sigma}(x(\lambda))\,\partial_\lambda x^\sigma(\lambda)}}\, d\lambda\,.
\end{equation*} The background	 Einstein tensor vanishes, $G_{\alpha\beta}=0$ and, to leading order in the perturbation induced by the light object, choosing $\lambda$ to be the background proper time, $\partial_\tau x^
\rho(\tau) g_{\rho\sigma}(x(\tau))\,\partial_\tau x^\sigma(\tau)=-1$, we arrive at \eqref{eq:linearizedEinsteinEquations} and \eqref{eq:stresspoint}}
\begin{equation}\label{eq:linearizedEinsteinEquations}
	\delta G_{\alpha\beta} = 8\pi G\, T_{\alpha\beta}\,,
\end{equation}
where
\begin{equation}\label{eq:stresspoint}	
	T^{\alpha\beta}(x) = \frac{\mu}{\sqrt{-g(x)}} \int \delta^{(4)}(x-x(\tau)) \,  \frac{d x^\alpha(\tau)}{d\tau} \frac{dx^\beta (\tau)}{d\tau}\, d\tau
\end{equation}
is the stress-energy tensor of the light particle and $\tau$ is the proper time.

For completeness, in addition to gravity, we also consider probe particles coupled to a electromagnetic (vector) field $A_\mu$ or to a scalar field $\varphi$. Denoting by $q_e$ and $q$ the electromagnetic and scalar charges, the fields emitted by such sources are given by 
\begin{equation}\label{eq:EOMs1s0}
	\nabla_\mu  F^{\mu\nu}
	=  - 4\pi J^\nu\,,
	\qquad
	\Box \varphi  = - 4\pi \rho\,,
\end{equation}
with $F_{\mu\nu} = \partial_\mu A_\nu - \partial_\nu A_\mu$ and
\begin{subequations}
	\begin{align}
		J^\alpha (x) 
		&= \frac{ q_e}{\sqrt{-g(x)}} \int \delta^{(4)}(x-x(\tau))
		\, \frac{d x^\alpha(\tau)}{d\tau}\, d\tau\,, \label{eq:J_expression}
		\\
		\rho (x) 
		&= \frac{q}{\sqrt{-g(x)}} \int \delta^{(4)}(x-x(\tau)) d\tau\,.
 \label{eq:rho_expression}
	\end{align}
\end{subequations}
For the electromagnetic case, it will be useful to define the Penrose scalars
\begin{equation}
	\mathsf{\Phi}_0 = F_{\mu\nu} \, \ell^\mu m^\nu\,,\qquad
	\mathsf{\Phi}_2 = F_{\mu\nu} \, \bar{m}^\mu n^\nu\,.
\end{equation}

\subsection{Teukolsky equation} \label{sec:teuk}

The wave equations for perturbations whose dynamics is described by \eqref{eq:linearizedEinsteinEquations} in the gravitational case ($s=\pm2$) and \eqref{eq:EOMs1s0} in the electromagnetic ($s=\pm1$) or scalar ($s=0$) cases can be cast in a unified form as follows \cite{Teukolsky:1972my},
\begin{equation}\label{eq:masterrt}
	\begin{split}
		&
		\frac{r^4}{\Delta}\,\partial_t^2 \psi - \frac{1}{(\sin\theta)^2}\,\partial_\phi^2\psi-\Delta^{-s}\partial_r(\Delta^{s+1}\partial_r\psi) - \frac{1}{\sin\theta}\,\partial_\theta(\sin\theta\partial_\theta\psi)
		\\
		&-2s \frac{i \cos\theta}{(\sin\theta)^2}\,\partial_\phi \psi
		-2s  \left(
		\frac{M r^2}{\Delta}-r
		\right)\partial_t \psi
		+
		(s^2(\cot\theta)^2-s)\psi
		=
		-\mathcal{T}\,.
	\end{split}
\end{equation}
Here, 
\begin{equation}
	\Delta(r) = r(r-2M)
\end{equation}
and the master variable $\psi$ and source $\mathcal{T}$ are given by \cite{Teukolsky:1972my}
\begin{equation}\label{eq:psisigma}
	\psi = \begin{cases}
		r^4 \mathsf{\Psi}_4 & \text{for }s=-2\\
		\mathsf{\Psi}_0 & \text{for }s=+2\\
		r^2 \mathsf{\Phi}_2 & \text{for }s=-1\\
		\mathsf{\Phi}_0 & \text{for }s=+1\\
		\varphi &\text{for }s=0
	\end{cases}
	\qquad
	\mathcal{T} = \begin{cases}
		4\pi G\, r^2 \mathcal{E}[T] & \text{for }s=-2\\
		4\pi G\, r^2 \tilde{\mathcal{E}}[T]& \text{for }s=+2\\
		4\pi r^2 \mathcal{E}[J]  & \text{for }s=-1\\
		4\pi r^2 \tilde{\mathcal{E}}[J]& \text{for }s=+1\\
		-4\pi r^2 \rho &\text{for }s=0
	\end{cases}
\end{equation}
in terms of differential operators $\mathcal{E}$, $\tilde{\mathcal{E}}$ detailed in Appendix~\ref{app:derivation_source_terms}.

Next, we introduce the mode decompositions
\begin{equation}\label{eq:Psi4modedecomposition}
	\psi = \int \frac{d\omega}{2\pi}\,\sum_{\ell,m}\,e^{-i\omega t} R_{\ell m}(\omega,r)\,Y_{s}^{\ell m}(\theta,\phi)
\end{equation}
and
\begin{equation}\label{eq:Tellm}
	\mathcal{T} = \int \frac{d\omega}{2\pi}\,\sum_{\ell,m}\,e^{-i\omega t} T_{\ell m}(\omega,r)\,Y_{s}^{\ell m}(\theta,\phi)\,,
\end{equation}
where $Y_{s}^{\ell m}(\theta,\phi) = e^{i m \phi } S_{s}^{\ell m}(\theta)$ are the spin-weighted spherical harmonics (see \eqref{eq:SWSH}) satisfying 
\begin{equation}
	\frac{\partial_\theta(\sin\theta \, \partial_\theta S_{s}^{\ell m})}{\sin\theta}
	+\left(
	(\ell-s)(\ell+s+1)
	-\frac{m^2}{(\sin\theta)^2}-\frac{2 m s \cos\theta}{(\sin\theta)^2}-s^2(\cot\theta)^2+s
	\right)
	S_{s}^{\ell m}=0\,.
\end{equation}
Using these variables, \eqref{eq:masterrt} reduces to the ordinary differential equation
\begin{align}\label{eq:TeukolskyEquation}
	&\frac{1}{\Delta(r)^s}\frac{d}{dr}\left[\Delta(r)^{s+1} \frac{d}{dr} R_{\ell m}(r)\right]
		\\
		&+
		\left(
		\frac{(r^2\omega)^2-2i s(r-M)r^2\omega}{\Delta(r)}+4i s \omega r -(\ell-s)(\ell+s+1)
		\right)R_{\ell m}(r)=T_{\ell m}(r)\,.
		\nonumber
\end{align}
Here and in the following, we shall omit the dependence of $R_{\ell m}$ and $T_{\ell m}$ on $\omega$ and, to avoid clutter, also the subscripts $\ell$ and $m$ when no confusion arises. 
Since we work in frequency domain at a fixed $\omega\neq 0$, we disregard $\delta$-function contributions localized at $\omega=0$, which translates to static terms in time domain. We will come back to this point below.

\subsection{Homogeneous equation} \label{sec:homogeneous}

Let us consider the homogeneous version of \eqref{eq:TeukolskyEquation},
\begin{align}\label{eq:TeukolskyHom}
	\frac{1}{\Delta^s}\frac{d}{dr}\left[\Delta^{s+1} \frac{d}{dr} R\right]
	+
	\left(
	\frac{(r^2\omega)^2-2i s(r-M)r^2\omega}{\Delta}+4i s \omega r -(\ell-s)(\ell+s+1)
	\right)R=0\,.
\end{align}
\eqref{eq:TeukolskyHom} is of the confluent Heun type, with two regular singularities at $r=0, 2M$ and an irregular singularity at infinity.
We introduce the dimensionless variables
\begin{equation}\label{eq:xyz}
	x = 4 i M \omega\,,\qquad
	y = 2i \omega r\,,\qquad
	z = \frac{2M}{r}
\end{equation}
such that
\begin{equation}
	R(r) = z^s (1-z)^{-\frac{s+1}{2}} \Psi(z) \Big|_{z=\frac{2M}{r}} 
\end{equation}
puts \eqref{eq:TeukolskyHom} in the Schr\"odinger-like form
\begin{equation}\label{eq:ODEpsi}
	\frac{d^2}{dz^2} \Psi  + Q\, \Psi =0
\end{equation}
with
\begin{equation}
		Q 
		=
		-\frac{x^2}{4 z^4}
		+
		\frac{x \left(s-\frac{x}{2}\right)}{z^3}
		+\frac{1-s^2}{4 (z-1) z}
		+\frac{1-(s+x)^2}{4 (z-1)^2 z}
		+
		\frac{4 \ell^2+4 \ell+x (3 x-2 s)}{4 (z-1) z^2}\,. \label{heun1}
\end{equation}
\eqref{eq:ODEpsi} is of the confluent Heun type with regular singularities at $z=1,\infty$ and an irregular one at $z=0$. 
Any equation of this type can be written as \eqref{eq:ODEpsi} with
\begin{equation}\label{eq:Qgauge}
	Q = -\frac{x^2}{4z^4}
	+ \frac{x m_3}{z^3}
	+
	\frac{1-(m_1-m_2)^2}{4(z-1)z}
	+
	\frac{1-(m_1+m_2)^2}{4(z-1)^2z}
	+
	\frac{\mathfrak{u}-\frac{1}{4}+\frac{x}{2}(m_1+m_2-1)}{(z-1)z^2}\,.
\end{equation}
It is straightforward to check that, by taking\footnote{Besides~\eqref{eq:dictionary}, there are other seven possible identifications between the gauge and gravity parameters that leave the structure of~\eqref{heun1}  unchanged. Indeed it is easy to check that $Q$ in ~\eqref{eq:Qgauge} is invariant under the following
replacements:  $m_1 \leftrightarrow m_2$, $(m_1,m_2,\mathfrak{u})\to (-m_1,-m_2,\mathfrak{u} + x(m_1+m_2))$ and $(x,m_3,\mathfrak{u})\to (-x,-m_3,\mathfrak{u} + x(m_1+m_2-1))$. To compare against \cite{Fucito:2024wlg,Consoli:2022eey}, one has to combine these transformations with  the CFT gauge theory dictionary,
 \be
 m_1=p_0-k_0 \,, \qquad m_2= -p_0-k_0 \,, \qquad m_3=c \,, \qquad x=-x_{\rm CFT}\,.
 \ee
 }
\begin{equation} \label{eq:dictionary}
	m_1 = \frac{x}{2}\,,\quad
	m_2 = s+\frac{x}{2}\,,\quad
	m_3 = s-\frac{x}{2}\,,\quad
	\mathfrak{u} = \left(\ell+\tfrac{1}{2}\right)^2+\frac{x}{2}-sx+\frac{x^2}{4}\,,
\end{equation}
\eqref{eq:Qgauge} reproduces~\eqref{heun1}.
\eqref{eq:ODEpsi} with $Q$ as in \eqref{eq:Qgauge}  describes the quantum Seiberg--Witten (SW) curve governing the dynamics of an $\mathcal{N}=2$ supersymmetric $SU(2)$ gauge theory with three fundamental hypermultiplets with masses $m_i$, coupling $x$ and Coulomb branch parameter $\mathfrak{u}$ living on a Nekrasov--Shatashvili curved background.

We introduce now two bases of solutions, $\{R^+_{H}(r),R^-_{H}(r)\}$ and 
$\{R^+_{\infty}(r),R^-_{\infty}(r)\}$, that are upgoing/ingoing at the horizon and at infinity respectively. They are specified by the asymptotics
\begin{subequations}\label{eq:asymptoticsRinfM}
\begin{align}
	\label{eq:asymptoticsRpmINF}
	R^{+}_{\infty} &\underset{r\to\infty\hspace{5pt}}{\sim} \left(\tfrac{2M}{r}\right)^{2s+1-2iM\omega} e^{i\omega r}\,,
	\qquad
	R^{-}_{\infty}\underset{r\to\infty\hspace{5pt}}{\sim}
	\left(\tfrac{2M}{r}\right)^{1+2iM\omega}\,e^{-i\omega r}\,, \\
		R^{+}_H &\underset{r\to2M}{\sim} 
	\left(1-\tfrac{2M}{r}\right)^{2iM\omega} e^{i\omega r}\,,
	\qquad 
	\hspace{6.4pt}
	R^{-}_H \underset{r\to2M}{\sim} 
	\left(1-\tfrac{2M}{r}\right)^{-s-2iM\omega} e^{-i\omega r}\,,
\end{align}
\end{subequations}
which can be found by solving the confluent Heun equation in the two limits.
A generic solution of \eqref{eq:TeukolskyHom} can then be written as
   \begin{equation}
	R =
	B^{\phantom{\infty}}_+ R^{+}_{\infty} + B^{\phantom{\infty}}_-  R^{-}_{\infty}  =D^{\phantom{H}}_+ R^{+}_H + D^{\phantom{H}}_- R^{-}_H \,,
\end{equation}
where $B_+$ ($B_-$)  and $D_+$ ($D_-$)  are the coefficients of the upgoing (ingoing) modes at infinity and horizon.    
  
For later convenience, we introduce the solutions  $\mathfrak{R}_\text{in}$, $\mathfrak{R}_\text{up}$ that satisfy \textit{purely} ingoing boundary conditions at the horizon and purely upgoing ones at infinity, $D^\text{in}_+=B_-^\text{up}=0$, and are normalized as follows, 
 \begin{subequations}\label{eq:defRinup}
\begin{align}
	\mathfrak{R}_\text{in} 
	&=
	D^\text{in}_- R^{-}_H=
	B^\text{in}_+ R^{+}_\infty + \frac{R^{-}_\infty}{4i M \omega}
	\,, \label{eq:defRin}\\
	\mathfrak{R}_\text{up}
	&=
	B^\text{up}_+ 
	R^{+}_\infty
	=
	\frac{R^+_H}{4iM\omega+s}
	+ 
	D^\text{up}_-
	R^-_H\,.\label{eq:defRup}
\end{align}
\end{subequations}
From these, we can construct the Wronskian rescaled by $\Delta(r)^{s+1}$,
\begin{equation}\label{eq:Wronskian}
	\mathcal{W} = \Delta(r)^{s+1} (\mathfrak{R}_\text{in}(r) \partial_r \mathfrak{R}_\text{up}(r)-\mathfrak{R}_\text{up}(r) \partial_r \mathfrak{R}_\text{in}(r))\,.
\end{equation}
This is independent of $r$ thanks to the fact that $\mathfrak{R}_\text{in/up}$ obey the homogeneous Teukolsky equation \eqref{eq:TeukolskyHom}. Therefore, we can calculate it either at infinity or at the horizon obtaining the same result:
\begin{equation}\label{eq:BBDD}
	\mathcal{W} = (2M)^{2s+1} B_+^\text{up} = (2M)^{2s+1} D_-^\text{in} \,.
\end{equation}
In this way we see that 
\begin{equation}\label{eq:asymptoticsW}
	\frac{\mathfrak{R}_\text{up}}{\mathcal{W}} \underset{r\to\infty}{\sim} \frac{e^{i\omega r_\ast}}{r^{2s+1}}
	\,,
	\qquad
	\frac{\mathfrak{R}_\text{in}}{\mathcal{W}} \underset{r\to2M}{\sim} \Delta(r)^{-s}\,\frac{e^{- i\omega r_\ast}}{2M}\,,
\end{equation}
where $r_\ast = r+2M \log\left(\frac{r}{2M}-1\right)$.

\subsection{Connection formulae and PM expansion}

Solutions of the confluent Heun equation are not known in analytic form, but can be systematically approximated in terms of hypergeometric functions order by order in a small-$x$ expansion \cite{Cipriani:2025ikx}, which corresponds to the $M\omega\to0$ limit.
 Here we briefly review this construction and refer to \cite{Cipriani:2025ikx} for further details. 
 We give results in terms of the general confluent Heun parameters $m_i$, $\mathfrak{u}$, so that they can be easily adapted to more general backgrounds like Kerr. 
 
 We first consider the so called \emph{exterior region}, where $M\ll r, \omega^{-1}$, which corresponds to small $x$ for generic $y$. There, a natural basis of solutions of \eqref{eq:TeukolskyHom} is given by (here and in the following $\alpha=\pm$)\footnote{These two values of $\alpha$ label the two elements of the basis in the exterior (or, later, interior) region we are considering in this section. They should not be confused with the indices $\pm$ that appear in \eqref{eq:asymptoticsRinfM}, which instead refer to the asymptotic behaviors at infinity and at the horizon.}
\begin{equation}
R_\alpha(r)  = \left.  e^{-{y\over 2}} (1-\ft{x}{y})^{  -{s\over 2}- {m_1+m_2\over 2}   } y^{-1-s+m_3}   G^0_{\alpha}(y)   \right|_{y=2{\rm i} \omega r}   \label{ralphag0}
\end{equation}
with
\begin{subequations}
	\begin{align}
G^0_\alpha(y) &=  P_0(y)  H^0_{\alpha}(y)  +\widehat{P}_0(y)  y H^0_{\alpha}{}'(y) \, ,\\
H^0_{\alpha}(y) &= y^{{1\over 2} -m_3-\alpha a}   {}_1 F_1 (\ft12 {-}m_3 {-} \alpha a,1{-}2 \alpha a;y )\,. \label{gh0}
	\end{align}
\end{subequations}
Here $P_0$, $\widehat{P}_0$ are infinite series in $x/y$ and $y$, but at the $k$th PM order, they become polynomials of order $k$ in these variables. In (\ref{gh0}) we have introduced the auxiliary parameter $a(\mathfrak{u})$ that characterizes the monodromy of solutions under rotations around $y=\infty$, which is the irregular singular point, 
\begin{equation}
		G_\alpha^0(e^{2i \pi} y) = e^{2i\pi (\frac{1}{2}-\alpha a-m_3)}\,G_\alpha^0(y)\,,
\end{equation}
and is defined implicitly by the inverse series
\be
\mathfrak{u}(a)=a^2+\sum_{i=1}^\infty \mathfrak{u}_i(a) x^i
\,.
\ee
In the gauge theory context, $a$ is the quantum SW period and $\mathfrak{u}(a)$ codifies the instanton prepotential 
\be
{\cal F}_{\rm inst} (a) =-\sum_{i=1}^\infty {\mathfrak{u}_i(a)\over i} x^i
\ee
through the quantum version of the Matone relation \cite{Matone:1995rx,Flume:2004rp}\footnote{ Alternatively, the series $a(\mathfrak{u})$ and $\mathfrak{u}(a)$ can be obtained by solving the infinite fraction equation \cite{Poghosyan:2020zzg,Consoli:2022eey}
\beaq
\label{fractionequality}
\frac{x M(a+1)}{P(a+1) -\frac{x M(a+2)}{P(a+2)-\, \cdots
}}
+\frac{x M(a)}{P(a-1)-\frac{x M(a-1)}{P(a-2)-\, \cdots
}}-P(a)=0
\eeaq
with
\be
P(a)=a^2{-}\mathfrak{u} {+}x\left( a{+}\ft12 {-}m_1{-} m_2{-}m_3  \right)  ~, \qquad\qquad    M(a)=  \prod_{i=1}^3\big( a- m_i-\tfrac{1}{2}\big)\, .
\ee
}
\begin{equation}
	\label{eq:Matone}
	\mathfrak{u}(a) = a^2 -x \, \partial_x  \mathcal{F}_\text{inst}(a)\,.
\end{equation}
  The coefficients $\mathfrak{u}_i$ and the ones that enter $P_0$, $\widehat{P}_0$ are determined recursively order by order in $x$, i.e.~as a PM expansion, by imposing the validity of the differential equation \cite{Cipriani:2025ikx}. 
The asymptotics of the exterior region solutions (\ref{ralphag0}) as $r\to\infty$  can be easily evaluated using the connection formulae of the hypergeometric functions $H^0_\alpha(y)$. 

To describe instead the \textit{interior region}, where $M, r \ll \omega^{-1}$,  we need a different picture of $R_\alpha$,  as an expansion in $x$, but keeping now $z$ generic. This can be obtained by
letting
\begin{equation}\label{eq:RalphaINT}
	R_\alpha(r) = \left. e^{-\frac{x}{2z}} 
	(
	1-z
	)^{-\frac{s}{2}-\frac{m_1+m_2}{2}}
	\left(\tfrac{x}{z}\right)^{-1-s+m_3}
	\frac{G^1_\alpha(z)}{g_\alpha(x)}  \right|_{z=\ft{2M}{r} }  
\end{equation}
with   
\begin{subequations}
	\begin{align}
    G^1_\alpha(z) &=     P_{1}(z) \, H^1_\alpha(z) {+} \widehat{P}_{1}(z) \,  z H^{1}_\alpha{}'(z)  \,,\\
    H^1_\alpha(z) &=  z^{-\frac{1}{2}+\alpha a+m_3} \,
   _2F_1(\ft12 {+} \alpha a {-} m_1,\ft12 {+} \alpha a {-} m_2;1+2\alpha 
   a;z).
  \label{g0}
  \end{align}
\end{subequations}
 Now  $P_{1}(z)$, $ \widehat{P}_{1}(z)$ are series in $x/z$ and $z$, and, at the $k$th PM order, they become polynomials of order $k$ in these variables.
 These solutions have the same monodromy as the $G_\alpha^0(y)$ around the irregular singular point $z=0$,  
\begin{equation}
	G_\alpha^1(e^{-2i\pi}\,z) = e^{2i\pi (\frac{1}{2}-\alpha a-m_3)}\, \,G_\alpha^1(z)\,.
\end{equation}
Thus, \eqref{ralphag0} and \eqref{eq:RalphaINT} define solutions of the same equation with the same monodromy around infinity, so they must be equal up to a $z$-independent
$g_\alpha(x)$.   This  normalization constant can be easily calculated by noting that $g_\alpha(x)= G_\alpha^1/G_\alpha^0$ and evaluating this ratio 
in the overlap of the two regions, the so called \textit{near zone},  $M\ll r \ll \omega^{-1}$, where $x$, $y$, $z$ are all small.  The solutions \eqref{eq:RalphaINT} can now be used to obtain the near-horizon asymptotics as $r\to2M$, i.e.~$z\to 1$, using again standard hypergeometric connection formulae.

The behaviour of $R_\alpha$ near the space time boundaries can be obtained using the standard hypergeometric connection formulae
\begin{subequations}\label{hyptrans}
\begin{align}
H^0_\alpha (y) & \underset{y\to \infty}{\sim} \sum_{\alpha' = \pm} B_{\alpha \alpha'} \, y^{-m_3(1+\alpha')} \, e^{\frac{y}{2}\left(1+\alpha'\right)} \, , \\
H^1_\alpha (z) & \underset{z \to 1}{\sim} \sum_{\alpha'=\pm} F_{\alpha \alpha'} \, (1-z)^{\frac{1+\alpha'}{2} (m_1+m_2)} \, . 
\end{align}
\end{subequations}
 with
\begin{subequations}
\begin{align}
B_{\alpha \alpha'} &=   e^{\frac{{\rm i} \pi (1{-}\alpha' )}{2}   ({1\over 2}{-}m_3{-}\alpha a )}  \frac{\Gamma (1{-}2  \alpha a )   }{\Gamma \left(\ft{1}{2}{-}\alpha a{-} \alpha'   m_3  \right)}\,, \\
F_{\alpha\alpha'}  &=  \frac{\Gamma(1{+}2\alpha a) \Gamma (-\alpha'(m_1+m_2)) }{\Gamma \left(\ft{1}{2}{+}\alpha a{-}\alpha'  m_1  \right)
	\Gamma \left(\ft{1}{2}{+}\alpha a{-}\alpha'  m_2  \right)}  \, ,
\end{align}
\end{subequations}
Performing the limits in (\ref{ralphag0}) and (\ref{eq:RalphaINT})  and using (\ref{hyptrans}) one finds \cite{Cipriani:2025ikx}
\begin{subequations}
	\begin{align}
R_\alpha(r) & \underset{r\to \infty}{\sim}   \sum_{\alpha' } B_{\alpha \alpha' } e^{ \alpha' {\rm i} \omega r}   (2{\rm i} \omega r)^{-1-s-\alpha' m_3  }       e^{ -\frac{ 1+\alpha'}{2} \partial_{m_3} {\cal F}_{\rm inst}} \, , \\
R_\alpha(r) & \underset{r\to 2M }{\sim}  \sum_{\alpha' } F_{\alpha \alpha'}   \tfrac{{h}_{\alpha'} }{ {g}_\alpha}  \, e^{-{\rm i} \omega r} (2{\rm i} \omega r)^{-1-s+m_3}    (1-\ft{2M}{r})^{ -{s\over 2} +(m_1+m_2) {\alpha'\over 2}} \, ,
\label{eq:conform}
\end{align}
\end{subequations}
with
\be
{{g}_\alpha} =c\,x^{\alpha a+m_3-{1\over 2} } e^{-{1\over 2} \alpha \partial_a {\cal F}_{\rm inst} } ~ , 
\qquad\qquad  
{{h}_{\alpha}} = c\, e^{-{1\over 2} ( \alpha  \partial_{m_1}  + \alpha  \partial_{m_2} +\partial_{m_3} ) {\cal F}_{\rm inst}  {+}x {\alpha{+}1\over 2}   } \,. \label{ids}
\ee
Here $c=c(x)$ is an $\alpha$-independent function of $x$ that cancels out in all ratios $g_\alpha/ g_\beta$, $h_\alpha/ g_\beta$ entering the computation of physical observables. 
The normalized ingoing and upgoing solutions defined in \eqref{eq:defRin}, \eqref{eq:defRup} can then be written as  
\begin{subequations}\label{eq:frakRinfrakRupGENERAL}
\begin{align}
\mathfrak{R}_{\rm in} (r) &=     C_{\rm in}   \left[  R_-(y)  -   \ft{ {{g}_+} F_{-+}}{ {{g}_-} F_{++}}  R_+(y)  \right]_{y=2{\rm i} \omega r}  \\
\mathfrak{R}_{\rm up} (r) &=        C_{\rm up}    \left[  R_+(y)   -  \ft{B_{+-}}{B_{--}}  R_-(y)  \right]_{y=2{\rm i} \omega r}  \label{rinup2}
\end{align}
\end{subequations}
with
\be
C_{\rm in}  =   {    x^{s-m_3 } \over B_{--} }
\left( 1-      \ft{ {{g}_+} F_{-+}  B_{+-} }{ {{g}_-} F_{++}B_{--}   } \right)^{-1} , \qquad   C_{\rm up}=   {x^{a+s+{1\over 2} }  e^{ {1\over 2} \left(\sum_i \partial_{m_i} {\cal F}_{\rm inst} -\partial_a {\cal F}_{\rm inst} \right) } \over   (m_1{+}m_2)    F_{+ +}  \left( 1-       \ft{ {{g}_+} F_{-+}  B_{+-} }{ {{g}_-} F_{++}B_{--}   } \right) }  \,.
\ee

\subsection{Inhomogeneous equation}

\label{sec:inhomogeneous}

The retarded solution of the inhomogeneous equation \eqref{eq:TeukolskyEquation} can be constructed 
by letting 
\begin{equation}\label{eq:solutionTeuk}
	R_{\ell m}( r) = \int_{2M}^\infty \mathcal{G}(r,r') \Delta(r')^{s}\, T_{\ell m}(r')\,dr'
\end{equation}
with\footnote{Note that $\mathcal{G}(r, r')$ is independent of the choices of normalization in \eqref{eq:defRin}, \eqref{eq:defRup} in view of \eqref{eq:Wronskian}.}
\begin{equation}\label{eq:propR}
	\mathcal{G}(r,r')= \frac{1}{\cal W}\left[
	\theta(r-r')
	\mathfrak{R}_\text{up}( r) 	\mathfrak{R}_\text{in}(r')
	+
	\theta(r'-r) 	\mathfrak{R}_\text{up}( r') 	\mathfrak{R}_\text{in}(r)
	\right],
\end{equation}
that is
\begin{equation}\label{eq:solutionTeukEXPL}
	\begin{split}
	R_{\ell m}(r) 
	&= \frac{	\mathfrak{R}_\text{up}(r)}{\cal W} \int_{2M}^r  	\mathfrak{R}_\text{in}(r') \Delta(r')^{s}\, T_{\ell m}(r')\,dr'\\
	&+ \frac{	\mathfrak{R}_\text{in}(r)}{\cal W} \int_{r}^\infty  	\mathfrak{R}_\text{up}(r') \Delta(r')^{s}\, T_{\ell m}(r')\,dr'\,.
	\end{split}
\end{equation}
For large values of $r$ and close to the horizon, one finds
\begin{subequations}
	\begin{align}
	\label{eq:asymptoticsinfinity}
	R_{\ell m}(r) 
	&\underset{r\to\infty}{\sim} \frac{e^{i\omega r_\ast}}{r^{2s+1}}\, Z_{\ell m, s}^\infty  \quad \text{for } s\le0 \,,
	\\
	\label{eq:lim2M}
	R_{\ell m}(r) 
	&\underset{r\to2M}{\sim} \frac{\Delta(r)^{-s}}{2M}\, e^{-i\omega r_\ast} Z_{\ell m, s}^H  \,,
\end{align}
\end{subequations}
with $r_\ast =  r + 2  M \log(\frac{r}{2M}-1)$ and
\begin{subequations}
	\begin{align}\label{eq:Zinfty}
		Z_{\ell m, s}^\infty 
		&= \int_{2M}^\infty \mathfrak{R}_\text{in}(r)\Delta(r)^{s}\, T_{\ell m}(r)\,dr\,,
		\\
		\label{eq:ZH}
		Z_{\ell m, s}^H  
		&= \int_{2M}^\infty \mathfrak{R}_\text{up}(r)\Delta(r)^{s} T_{\ell m}(r)\,dr\,,
	\end{align}
\end{subequations}
as we now discuss.
To study the near-horizon behavior of \eqref{eq:solutionTeukEXPL}, let us rewrite it in the equivalent form
\begin{align}\label{eq:solutionTeukEXPL3}
	&R_{\ell m}(r) 
	= \frac{\mathfrak{R}_\text{in}(r)}{\mathcal{W}}  \, Z_{\ell m,s}^H \\
	&
	+ \frac{\mathfrak{R}_\text{up}(r)}{\mathcal{W}} \int_{2M}^{r}	\mathfrak{R}_\text{in}(r') \Delta(r')^{s}\, T_{\ell m}(r')\,dr'
	-
	\frac{	\mathfrak{R}_\text{in}(r)}{\mathcal{W}} \int_{2M}^r	\mathfrak{R}_\text{up}(r') \Delta(r')^{s}\, T_{\ell m}(r')\,dr'\,.
	\nonumber
\end{align} 
For scatterings at a fixed impact parameter, which is the case of interest here, the source is localised on the particle's trajectory and is always far from the horizon.  Therefore, for  any $r$ smaller than the radius of closest approach, the second line of \eqref{eq:solutionTeukEXPL3} vanishes and one finds 
\begin{equation}
	R_{\ell m}(r) =  \frac{\mathfrak{R}_\text{in}(r)}{\mathcal{W}} \, Z_{\ell m,s}^H \,,
\end{equation}	
leading to \eqref{eq:lim2M} as $r\to2M$.
	At large distances, even for very large $r$, there  is always a part of the trajectory for which $r'>r$. Rewriting \eqref{eq:solutionTeukEXPL} as
\begin{align}\label{eq:solutionTeukEXPL2}
		&R_{\ell m}(r) 
		= \frac{\mathfrak{R}_\text{up}(r)}{\mathcal{W}}  \, Z_{\ell m,s}^\infty \\
		&
		+ \frac{\mathfrak{R}_\text{in}(r)}{\mathcal{W}} \int_{r}^\infty  	\mathfrak{R}_\text{up}(r') \Delta(r')^{s}\, T_{\ell m}(r')\,dr'
		-
		\frac{	\mathfrak{R}_\text{up}(r)}{\mathcal{W}} \int_{r}^\infty  	\mathfrak{R}_\text{in}(r') \Delta(r')^{s}\, T_{\ell m}(r')\,dr'\,,
		\nonumber
\end{align} 
both terms in the second line can be in principle nontrivial even for large $r$. However, by \eqref{eq:defRinup}, \eqref{eq:asymptoticsRpmINF}, they are clearly subdominant for $s\le0$, so that we can focus on the first line of \eqref{eq:solutionTeukEXPL2}, leading to \eqref{eq:asymptoticsinfinity} as $r\to\infty$.
Instead, the terms in the second line of \eqref{eq:solutionTeukEXPL2} can be relevant for $s=1,2$ and it would be interesting to evaluate them explicitly. We will come back to their impact on the field at null infinity in the next section.

\subsection{Waveforms} \label{sec:waveforms}

In this section, we study the behavior of the Penrose scalars dictated by the Teukolsky equation at future null infinity and close to the horizon. We then connect them to the corresponding waveforms.

\subsubsection*{Asymptotic behavior at infinity}

Away from the black hole, we introduce the retarded time $u=t-r_\ast$ and work in retarded coordinates $(u,r,\theta,\phi)$. The  form of the Schwarzschild metric  in these coordinates is  given in \eqref{eq:SchwRet}.  
Plugging the asymptotics \eqref{eq:asymptoticsinfinity} into the mode expansion \eqref{eq:Psi4modedecomposition}, we deduce the asymptotics for the relevant (Penrose) scalars at infinity. 
For large $r$ at fixed $u$, $\theta$, $\phi$, we find
\begin{subequations}\label{eq:nearinfinity}
	\begin{align}
		\mathsf{\Psi}_4 &\underset{r\to\infty}{\sim} \frac{1}{r} \int \frac{d\omega}{2\pi} \sum_{\ell m} e^{-i\omega u} Z^\infty_{\ell m,-2}(\omega) Y_{-2}^{\ell m}(\theta, \phi)\,,
		\\
		\label{eq:Psi0peeling}
		\mathsf{\Psi}_0 &\underset{r\to\infty}{\sim} \frac{1}{r^5} \int \frac{d\omega}{2\pi} \sum_{\ell m} e^{-i\omega u} Z^\infty_{\ell m,+2}(\omega) Y_{+2}^{\ell m}(\theta, \phi)\,,
		\\
		\mathsf{\Phi}_2 &\underset{r\to\infty}{\sim} \frac{1}{r} \int \frac{d\omega}{2\pi} \sum_{\ell m} e^{-i\omega u} Z^\infty_{\ell m,-1}(\omega) Y_{-1}^{\ell m}(\theta, \phi)\,,
		\\
		\label{eq:Phi0peeling}
		\mathsf{\Phi}_0 &\underset{r\to\infty}{\sim} \frac{1}{r^3} \int \frac{d\omega}{2\pi} \sum_{\ell m} e^{-i\omega u} Z ^\infty_{\ell m,+1}(\omega) Y_{+1}^{\ell m}(\theta, \phi)\,,
		\\
		\varphi & \underset{r\to\infty}{\sim} \frac{1}{r} \int \frac{d\omega}{2\pi} \sum_{\ell m} e^{-i\omega u} Z^\infty_{\ell m,0}(\omega) Y_{0}^{\ell m}(\theta, \phi)\,.
	\end{align}
\end{subequations}
In writing the asymptotic behaviors \eqref{eq:Psi0peeling}, \eqref{eq:Phi0peeling}, we have neglected the terms in the second line of \eqref{eq:solutionTeukEXPL2}, thus employing a behavior formally identical to \eqref{eq:asymptoticsinfinity} also for $s=1,2$. Such additional terms arise from the portion of the trajectory $r'>r$, where $r$ is the distance between the source and the observer. Therefore, when inserted in the Fourier transform \eqref{eq:Psi4modedecomposition} and evaluated as $r\to\infty$ for fixed $u=t-r_\ast$, they will be localised at frequencies $\omega\sim\mathcal{O}(1/r)$. Since the distance $r$ is much larger than any length scale defining the binary, they are effectively zero-frequency $\delta(\omega)$ terms and are not captured by our approach.

To compare with the metric perturbation, we impose the retarded Bondi gauge \cite{Bondi:1962px,Sachs:1962wk,Sachs:1962zza,Strominger:2013jfa,Veneziano:2025ecv},
\begin{equation}\label{eq:BondiGauge}
	h_{rr}(u,r,\theta,\phi)=h_{rA}(u,r,\theta,\phi)=0\,,\qquad \gamma^{AB} h_{AB}(u,r,\theta,\phi)=0\,,
\end{equation}
where $A,B$ can take values $\theta$, $\phi$ and $\gamma_{AB}$ is the metric on the round sphere \eqref{eq:roundspheremetric}. We also assume the standard falloff $h_{AB}(u,r,\theta,\phi)  \sim \mathcal{O}(r)$ as $r\to\infty$,
while the other nonvanishing components are at most $\mathcal{O}(r^0)$. For the vector case, we similarly impose the retarded radial gauge \cite{Campiglia:2015qka,Strominger:2017zoo},
\begin{equation}
	A_r(u,r,\theta,\phi) = 0
\end{equation} 
and the large-$r$ falloff $A_A(u,r,\theta,\phi)  {\sim} \mathcal{O}(r^0)$ as $r\to\infty$
while $A_u(u,r,\theta,\phi) \sim \mathcal{O}(r^{-1})$ in the same limit.
We then find the following asymptotic relations between the Penrose scalars and the gauge fields as $r\to\infty$ for fixed $u$, $\theta$, $\phi$,
\begin{equation}\label{eq:asymptinftyPsi4Phi2}
	\mathsf{\Psi}_4 
	\underset{r\to\infty}{\sim}  - \frac{1}{2}\,\partial^2_u h
	\,,\qquad
	\mathsf{\Phi}_2 
	\underset{r\to\infty}{\sim} - \partial_u A\,,
\end{equation}
where
\begin{equation}
	h
	=
	\bar{m}^\mu h_{\mu\nu} \bar{m}^\nu\,,
	\qquad
	A
	=
	\bar{m}^\mu A_\mu\,.
\end{equation}
Let us recall that the complex variable $h$ is linked to the standard real polarizations of the gravitational waves $h_+$, $h_{\times}$ by $h = h_+-ih_\times$.
Let us introduce  the multipolar waveforms at infinity $W_{\ell m,s}^\infty$ 
\begin{subequations}
	\begin{align}\label{eq:modeh}
		h &\underset{r\to\infty}{\sim} \frac{4G}{r}\int \frac{d\omega}{2\pi} \sum_{\ell m} e^{-i\omega u} W^{\infty}_{\ell m,-2}(\omega) Y_{-2}^{\ell m}(\theta, \phi)\,,
		\\
		A &\underset{r\to\infty}{\sim}   \frac{1}{r} \int \frac{d\omega}{2\pi} \sum_{\ell m} e^{-i\omega u} W_{\ell m,-1}^{\infty} (\omega) Y_{-1}^{\ell m}(\theta,\phi)\,,
		\\
		\varphi
		&\underset{r\to\infty}{\sim} \frac{1}{r} \int \frac{d\omega}{2\pi} \sum_{\ell m} e^{-i\omega u} W_{\ell m,0}^\infty(\omega) Y_{0}^{\ell m}(\theta,\phi)\,,
	\end{align}
\end{subequations}
that can be found from the knowledge of $Z_{\ell m,s}^\infty$ defined in \eqref{eq:Zinfty} via \eqref{eq:asymptinftyPsi4Phi2} and \eqref{eq:nearinfinity}, in particular 
\begin{align}\label{eq:WinfZinf}
		W_{\ell m,-2}^\infty = \frac{1}{2G\omega^2}\,Z^\infty_{\ell m,-2}
		\,,
		\qquad
		W_{\ell m,-1}^{\infty} = \frac{1}{i \omega} \, Z_{\ell m,-1}^{\infty}
		\,,
		\qquad
		W_{\ell m,0}^{\infty} = Z_{\ell m,0}^{\infty}
		\,.
\end{align}
In particular, the first relation in \eqref{eq:WinfZinf} agrees with \cite{Fucito:2024wlg}.
We conclude this section with some further comments on the falloff of our general expressions \eqref{eq:nearinfinity}. Examples of asymptotic solutions to the Einstein equations violating the leading-order peeling condition $\mathsf{\Psi}_{2-s} \sim \mathcal{O}(r^{-s-3})$  are discussed in the literature, see for instance~\cite{Damour:1985cm,Christodoulou:1986du,Winicour1985} and the recent account~\cite{Geiller:2024ryw} focusing on the case of gravitational scattering. In particular, it was pointed out that in the latter case $\mathsf{\Psi}_0$ can decay as $\mathcal{O}(r^{-4})$, which is slower than the result quoted in~\eqref{eq:Psi0peeling}. We notice that  the peeling violating term in these papers is static, i.e.~$u$-independent. In frequency domain, this translates into $\delta(\omega)$ contributions,  which are not included in our analysis.

The property that  the $\mathcal{O}(r^{-4})$ term in $\mathsf{\Psi}_0$ is static is in agreement with previous results stating that the first correction to the shear tensor, i.e.~$h^{(0)}_{AB}$ in\footnote{Notice that \eqref{eq:hexp} corresponds to the falloff condition we discussed just below \eqref{eq:BondiGauge}.}
\begin{equation}
    \label{eq:hexp}
  h_{AB}(u,r,\theta,\phi)=r \, h^{(-1)}_{AB}(u,\theta,\phi)+ h^{(0)}_{AB}(u,\theta,\phi)+\cdots \,  
\end{equation}
is static, see (4.48) of~\cite{Barnich:2010eb} ($\partial_u h^{(0)}_{AB} = 0$ in the notation above). This result determines the behavior of $\mathsf{\Psi}_0$ thanks to the following simple relation with the metric fluctuation,
\begin{equation}\label{eq:Psi0hExact}
	\mathsf{\Psi}_0 =-\frac{1}{2r^4}\,(2-2r\partial_r + r^2\partial_r^2)  \left(h_{\theta\theta}+i (\sin\theta)^{-1} h_{\theta\phi}\right),
\end{equation}
leading to
\begin{equation}\label{eq:peelingviolation0}
	\mathsf{\Psi}_0 \sim - \frac{1}{r^4}\left(h^{(0)}_{\theta\theta}+i (\sin\theta)^{-1} h^{(0)}_{\theta\phi}\right),
\end{equation}
which then inherits the same $u$-dependence as $h^{(0)}_{AB}$. However,  very recently in \cite{DeAngelis:2025vlf}, a new and possibly non-static  peeling violating contribution to $\mathsf{\Psi}_0$  at order $\mathcal{O}(r^{-4})$  was found in the 
two-body scattering. We do not find such type of contributions in our approach. Peeling violations have been studied in even more general setups and at subleading orders in the large $r$ expansion, see for instance~\cite{ValienteKroon:1998vn} and recently~\cite{Geiller:2024ryw}. These studies indicate that asymptotic flatness and Bianchi identities require that all the terms of order $\mathcal{O}(r^{-4})$ are constants of motion~\cite{ValienteKroon:1998vn}.\footnote{RR would like to thank Juan Valiente-Kroon for discussions and explanations on this point.} This supports  our results above. It would be interesting to understand how the findings of \cite{DeAngelis:2025vlf} can consistently fit in this picture.

\subsubsection*{Behavior at the horizon}

Close to the black hole, it is more convenient to introduce the advanced time $v=t+r_\ast$ and to work in the advanced coordinates $(v,r,\theta,\phi)$.  The form of the Schwarzschild metric in these coordinates is given in  \eqref{eq:SchwAdv}. We find the following near-horizon asymptotics for the gravitational Penrose scalars
\begin{subequations}
	\begin{align}
		r^4\mathsf{\Psi}_4 
		&\underset{r\to2M}{\sim} \frac{\Delta(r)^2}{2M} \int \frac{d\omega}{2\pi} \sum_{\ell m} e^{-i \omega v} Z_{\ell m,-2}^H(\omega) Y_{-2}^{\ell m}(\theta,\phi)
		\,,\label{eq:nearhorizon_psi4}
		\\
		\mathsf{\Psi}_0 &\underset{r\to2M}{\sim} \frac{\Delta(r)^{-2}}{2M} 
		\int \frac{d\omega}{2\pi} \sum_{\ell m} e^{-i \omega v} Z_{\ell m,+2}^H(\omega) Y_{+2}^{\ell m}(\theta,\phi)\,.\label{eq:nearhorizon_psi0}
	\end{align}
\end{subequations}
There is a relation between the  coefficients $Z^H_{\ell m,-2}$ and ${Z}^H_{\ell m,+2}$ given by \cite{Teukolsky:1974yv,Tagoshi:1997jy,Pound:2021qin}\footnote{To facilitate comparison with \cite{Teukolsky:1974yv}, we recall that their $Z_\text{hole}$ is our $Z_{\ell m,-2}^{H}(\omega)/(2M)$ and
	their $Y_\text{hole}$ is our ${Z}_{\ell m,+2}^{H}(\omega)/(2M)$.  }
\begin{equation}\label{eq:magicYZ}
	C \,{Z}_{\ell m,+2}^{H}(\omega)
	=
	16 i \omega (2 M)^5  (1-2 i M \omega ) \left(1+16 M^2 \omega ^2\right)
	Z_{\ell m,-2}^{H}(\omega)
\end{equation}
with 
\begin{equation}
	C = \ell (\ell^2-1)(\ell+2) +  (-1)^{\ell + m}\,12 i M \omega\,.
\end{equation}
Similarly, for the electromagnetic scalars,
\begin{subequations}
	\begin{align}
		r^2 \mathsf{\Phi}_2 &\underset{r\to2M}{\sim} \frac{\Delta(r)}{2M} \int \frac{d\omega}{2\pi} \sum_{\ell m,-1} e^{-i\omega v} Z_{\ell m,-1}^H(\omega) Y_{-1}^{\ell m}(\theta,\phi)\,, \label{eq:nearhorizon_phi2}
		\\
		\mathsf{\Phi}_0 &\underset{r\to2M}{\sim} \frac{\Delta(r)^{-1}}{2M} \int \frac{d\omega}{2\pi} \sum_{\ell m} e^{-i\omega v} Z_{\ell m,+1}^H(\omega) Y_{+1}^{\ell m}(\theta,\phi)\,, \label{eq:nearhorizon_phi0}
	\end{align}
\end{subequations}
with \cite{Teukolsky:1974yv}
\begin{equation}\label{eq:magicYZ1}
	B \,{Z}_{\ell m,+1}^H(\omega) = - 4 i \omega (2M)^3(1-4i M \omega) Z_{\ell m,-1}^H(\omega)\,,
	\qquad
	B = \ell (\ell+1)\,,
\end{equation}
and
\begin{equation}
	\varphi \underset{r\to2M}{\sim} \frac{1}{2M} \int \frac{d\omega}{2\pi} \sum_{\ell m} e^{-i\omega v} Z_{\ell m,0}^H(\omega) Y_{0}^{\ell m}(\theta,\phi)\,.
\end{equation}

Moving to the metric perturbation, we impose the advanced Bondi gauge,
\begin{equation}\label{eq:BondiGauge2}
	h_{rr}(v,r,\theta,\phi)=h_{rA}(v,r,\theta,\phi)=0\,,\qquad \gamma^{AB} h_{AB}(v,r,\theta,\phi)=0\,.
\end{equation}
We also assume that, as $r\to2M$, the  $h_{AB}(v,r,\theta,\phi)$ components are finite
%\begin{equation}
%	h_{AB}(v,r,\theta,\phi) \underset{r\to2M}{\sim} 	\mathcal{O}((r-2M)^0)
%\end{equation} 
while $h_{vv}(v,r,\theta,\phi)$ and $h_{vA}(v,r,\theta,\phi)$ vanish at the horizon \cite{Donnay:2015abr}.\footnote{Note that~\cite{Donnay:2015abr} works in a different near-horizon gauge.}
For the vector case, we impose the advanced radial gauge
\begin{equation}
	A_r(v,r,\theta,\phi) = 0
\end{equation}
and similarly assume that $A_A(v,r,\theta,\phi)$ is finite as $r\to2M$ and that $A_v(v,r,\theta,\phi)$ vanishes close to the horizon.
Doing so, we find
\begin{equation}\label{eq:asymptPsi0Phi0H}
	\Delta(r)^2 \mathsf{\Psi}_0 \sim (2M)^3 (1-4M \partial_v)\partial_v\bar{h}\,,
	\qquad
	\Delta(r) \mathsf{\Phi}_0 \sim 2(2M)^2\,\partial_v \bar{A}\,,
\end{equation}
in terms of 
\begin{equation}
	\bar{h}=m^\mu h_{\mu\nu} m^\nu\,,\qquad
	\bar{A}=m^\mu A_\mu\,.
\end{equation}
The first relation in \eqref{eq:asymptPsi0Phi0H} agrees with~\cite{Poisson:2004cw}.
We introduce the mode decompositions of $\bar{h},\bar{A}$ and $\varphi$ and the ``waveforms at the horizon'' $W_{\ell m,s}^H$ by
\begin{subequations}
	\begin{align}\label{eq:modebar}
		\bar{h} &\underset{r\to2M}{\sim}
		\frac{4G}{2M}
		\int \frac{d\omega}{2\pi}\sum_{\ell m} e^{-i \omega v} W_{\ell m,+2}^H(\omega) Y^{\ell m}_{+2}(\theta,\phi)\,,\\
		\bar{A} &\underset{r\to2M}{\sim} \frac{1}{2M} \int \frac{d\omega}{2\pi}\,\sum_{\ell m} e^{-i\omega v} W_{\ell m,+1}^{H} (\omega) Y_{+1}^{\ell m}(\theta,\phi)\,,
		\\
		\varphi 
		&\underset{r\to2M}{\sim} \frac{1}{2M} \int \frac{d\omega}{2\pi} \sum_{\ell m} e^{-i\omega v} W_{\ell m,0}^H(\omega) Y_{0}^{\ell m}(\theta,\phi)\,.
	\end{align}
\end{subequations}
Using  \eqref{eq:nearhorizon_psi0}, \eqref{eq:nearhorizon_phi0} and the relations \eqref{eq:asymptPsi0Phi0H}, we thus find that they are fixed in terms of the coefficients ${Z}_{\ell m,s}^H$ given by \eqref{eq:ZH} according to
\begin{equation}\label{eq:WHtildeZ}
		{Z}^H_{\ell m,+2} = -4iG \omega (2M)^3 
		(1+4 i M\omega)  W_{\ell m,+2}^H\,,
		\qquad
		{Z}_{\ell m,+1}^H = 2(2M)^2(-i \omega) W_{\ell m,+1}^{H}\,.
\end{equation}
To find the relation with $Z_{\ell m,-|s|}^H$, we use \eqref{eq:magicYZ}, \eqref{eq:magicYZ1} obtaining
\begin{subequations}
	\begin{align}
		W_{\ell m,+2}^{H}
		&=
		-
		\frac{4}{C G} \,(2 M)^2  (1-2 i M \omega )(1-4 i M\omega)
		Z_{\ell m,-2}^{H}\,, \label{eq:WHZa}
		\\
		W_{\ell m,+1}^H 
		&=  \frac{4M}{B}(1-4i M \omega) Z_{\ell m,-1}^H\,. \label{eq:WHZb}
	\end{align}
\end{subequations}
Finally,
\begin{equation}\label{eq:WHZ0}
	W_{\ell m,0}^H = Z_{\ell m,0}^{H}\,.
\end{equation}

\subsection{The source term} \label{sec:source_term}
To calculate $Z_{\ell m,s}^{\infty/H}(\omega)$ from \eqref{eq:Zinfty}, \eqref{eq:ZH}, we need to use the explicit expression of $T_{\ell m}$ defined by \eqref{eq:Tellm}, \eqref{eq:psisigma}. We focus on point particles moving in the $\theta=\frac{\pi}{2}$ plane and perform the derivation explicitly in Appendix~\ref{app:derivation_source_terms}. Here we display the final results. We write  
\begin{equation}\label{eq:ZbL}
	Z_{\ell m,s}^{\infty/H} 
	=
	\int  dt \,e^{i\omega t - i m \phi(t)}
	\sum_{i=0}^2 b^i_{\ell m}\,\mathcal{L}_{i} [\mathfrak{R}_{\text{in}/\text{up}}(r(t))]\,,
\end{equation}
where  $b^i_{\ell m}$, $\mathcal{L}_{i}$ depend on $s$. 

\begin{itemize}

\item $s=- 2:$

\begin{subequations}  \label{eq:b_sm2}
	\begin{align}
		b^{0}_{\ell m}
		&=
		\pi \sqrt{(\ell^2-1)\ell(\ell+2)}\,Y_{0}^{\ell m}(\tfrac{\pi}{2},0)\,,
		\\
		b^{1}_{\ell m}
		&=
		2\pi \sqrt{(\ell-1)(\ell+2)}\,Y_{-1}^{\ell m}(\tfrac{\pi}{2},0)\,,
		\\
		b^{2}_{\ell m}
		&= 2\pi Y_{-2}^{\ell m}(\tfrac{\pi}{2},0)
	\end{align}
\end{subequations}
and 
\begin{subequations}\label{eq:L_sm2}
	\begin{align}
		\mathcal{L}_{0} &=G \frac{E}{r^2 f}\left(
		1+ \frac{\dot{r}}{f}
		\right)^2,
		\\
		\mathcal{L}_{1} &=G \frac{iL}{r^3}
		\left(
		1+ \frac{\dot{r}}{f}
		\right)\left(2-r\partial_r+\frac{i\omega r}{f}\right),
		\\
		\mathcal{L}_{2} &=G \frac{L^2 f}{Er^4}
		\left[
		\frac{1}{f^2}
		\left(
		i\omega M+\frac{\omega^2 r^2}{2}
		-i\omega r
		\right)
		+
		\left(
		1+\frac{i\omega r}{f}
		\right)
		r\partial_r
		-\frac{1}{2}\,r^2 \partial_r^2
		\right]\,,
	\end{align}
\end{subequations}
where $\dot{r}=\frac{d r}{dt}$.
Thus, we recover the expression of \cite{Fucito:2024wlg}.

\item $s= 2:$

\begin{subequations} \label{eq:b_sp2}
	\begin{align}
		b^{0}_{\ell m}
		&=
		\pi \sqrt{(\ell^2-1)\ell(\ell+2)}\,Y_{0}^{\ell m}(\tfrac{\pi}{2},0)\,,
		\\
		b^{1}_{\ell m}
		&=
		2\pi \sqrt{(\ell-1)(\ell+2)}\,Y_{1}^{\ell m}(\tfrac{\pi}{2},0)\,,
		\\
		b^{2}_{\ell m}
		&= 2\pi Y_{2}^{\ell m}(\tfrac{\pi}{2},0)
	\end{align}
\end{subequations}
and 
  \begin{subequations} \label{eq:L_sp2}
	\begin{align} 
		\mathcal{L}_{0} &= 4 G r^2 f E \left(
		1- \frac{\dot{r}}{f}
		\right)^2,
		\\
		\mathcal{L}_{1} &=G  iL
		\left(
		1 - \frac{\dot{r}}{f}
		\right)\left(
		4 r^2 f^2 \left(
		\partial_r + \frac{i\omega}{f}
		\right)
		+8rf
		\right),
		\\
	 \mathcal{L}_{2} &= - G \frac{2L^2 f}{E}
		\left[
		r^2f^2 \left(
		\partial^2_r + \frac{2 i\omega}{f} \partial_r -  \frac{2 i M \omega}{f^2 r^2} - \frac{\omega^2}{f^2}
		\right) 
		\right. \notag
		\\
		&
		 \left.+2(3r-2M)f\left(
		\partial_r + \frac{i\omega}{f}
		\right) 
		 +4\left(1-\frac{2M^2}{r^2}\right)
		\right].
	\end{align}
\end{subequations}

\item $s=- 1:$

\begin{align} \label{eq:b_sm1}
	b^{0}_{\ell m}
	=
	4\pi \sqrt{\ell(\ell+1)}\,Y_{0}^{\ell m}(\tfrac{\pi}{2},0)\,,
	\qquad
	b^{1}_{\ell m}
	=
	4\pi Y_{-1}^{\ell m}(\tfrac{\pi}{2},0)\,, \qquad  b^{2}_{\ell m}=0
\end{align}
and 
\begin{align} \label{eq:L_sm1}
	\mathcal{L}_{0} = \frac{q_e }{2\sqrt2 r}\left(
	1+ \frac{\dot{r}}{f}
	\right),
	\qquad
	\mathcal{L}_{1} = \frac{q_e  L}{E}\,\frac{i f}{2\sqrt2 r^2}
	\left(-r\partial_r+\frac{i\omega r}{f}+1\right).
\end{align}

\item $s=+1:$

\begin{align} \label{eq:b_sp1}
	b^{0}_{\ell m}
	=
	4\pi \sqrt{\ell(\ell+1)}\,Y_{0}^{\ell m}(\tfrac{\pi}{2},0)\,,
	\qquad
	b^{1}_{\ell m}
	=
	4\pi Y_{1}^{\ell m}(\tfrac{\pi}{2},0)\,, \qquad  b^{2}_{\ell m}=0
\end{align}
and 
\begin{equation} \label{eq:L_sp1}
	\mathcal{L}_0 =q_e \frac{rf}{\sqrt{2}}\left(
	1-\frac{\dot{r}}{f}
	\right),\qquad
	\mathcal{L}_1 = q_e\frac{L}{E} \frac{if^2}{\sqrt{2}}
	\left(
	 r\partial_r + {i \omega r\over f}  +{1\over f} 
	\right).
\end{equation}

\item $s=0:$

\be \label{eq:b_s0}
 b^{0}_{\ell m}
	=
	4\pi \,Y_{0}^{\ell m}(\tfrac{\pi}{2},0)\,, \qquad  b^{1}_{\ell m}=b^{2}_{\ell m}=0
 \ee
 and
\begin{equation} \label{eq:L_s0}
	 {\cal L}_0=	-
	 q \mu 
	\frac{f}{E}  \,.
\end{equation}

\end{itemize}

\subsection{Radiated and absorbed energy and angular momentum} \label{sec:fluxes}

The energy and angular momentum  absorbed by the black hole are given by the formulae~\cite{Poisson:2004cw}\footnote{
While \cite{Teukolsky:1974yv} lays out the frequency-domain description of horizon amplitudes and absorption, \cite{Poisson:2004cw} provides a unified treatment of absorbed energy and \emph{angular momentum}, so we focus on the latter.}
\begin{subequations}\label{eq:absEJbasic}
	\begin{align}
		E_\text{abs} &= \frac{1}{32\pi G} \lim_{r\to2M} \int dv\oint r^2 d\Omega \, \partial_v h_{AB} \partial_v h^{AB}\,,
		\\
		J_\text{abs} &= -\frac{1}{32\pi G}  \lim_{r\to2M} \int dv \oint r^2d\Omega \, \partial_v h_{AB} \partial_\phi  h^{AB}\,.
	\end{align}
\end{subequations}
Formally, identical expressions describe the energy and angular momentum carried to infinity by the gravitational field after replacing the limit by $r\to \infty$ and $v$ by the retarded time $u$ \cite{Blanchet:2013haa,Damour:2020tta,Heissenberg:2024umh}. 
Here the indices $A$, $B$ are raised using $g^{AB} = r^{-2}\gamma^{AB}$ and we have indicated $d \Omega = \sin \theta \, d\theta \, d\phi$. The fluxes  \eqref{eq:absEJbasic}, and analogous ones for vector and scalar fields discussed below, can be derived by applying the Noether theorem to the actions for linearized perturbations on the Schwarzschild background \eqref{eq:Schwarzschild}. See Appendix~\ref{app:fluxes} for this derivation.

Going to the frequency domain (see e.g.~the identities in (A.2) of \cite{Heissenberg:2024umh}) and performing the mode decomposition as in \eqref{eq:modebar}, we find
\begin{equation} \label{eq:absEJgr}
	E_\text{abs} = \frac{G}{\pi^2} \int_0^\infty d\omega \, \omega^2 \sum_{\ell m} \left|{W}^H_{\ell m,+2}\right|^2\,,
	\qquad
	J_\text{abs} = \frac{G}{\pi^2} \int_0^\infty d\omega \, \omega \sum_{\ell m} m \left|{W}^H_{\ell m,+2}\right|^2\,.
\end{equation}
Here we used the fact that the two helicities give the same contribution.\footnote{The reality of the position-space field imposes $W^{H}_{\ell m,- s}(\omega) = (-1)^{s+m} \big[W^{H}_{\ell(-m),+ s}(-\omega)\big]^\ast$, where $\ast$ stands for complex conjugation, and, for the results in Section~\ref{sec:computation_all}, this is equivalent to $W^{H}_{\ell m,- s}(\omega) = (-1)^{s+\ell+m} W^{H}_{\ell m, +s}(\omega)$ as expected by parity. So, $|W_{\ell m,-2}^H|^2=|W_{\ell m,+2}^H|^2$. 
Identical considerations hold for the waveform at infinity \cite{Fucito:2024wlg}. \label{footnote:Parity}} 
The same formulae also describe the radiated energy and angular momentum after replacing ${W}^H_{\ell m,+2}$ by  ${W}^\infty_{\ell m,-2}$. \\
Similar results are obtained for the vector and scalar cases
\begin{align}
	E^{\rm vector}_\text{abs} &= \frac{2}{\pi} \int_0^\infty d\omega\,\omega^2\sum_{\ell m} \left| W^H_{\ell m,+1}\right|^2\,,
	\qquad
	J^{\rm vector}_\text{abs} = \frac{2}{\pi} \int_0^\infty d\omega\,\omega\sum_{\ell m} m \left| W^H_{\ell m,+1}\right|^2\,, \label{eq:absEJem} \\
	E^{\rm scalar}_\text{abs} &= \frac{1}{\pi} \int_0^\infty d\omega\,\omega^2\sum_{\ell m} \left| W^H_{\ell m,0}\right|^2\,,
	\qquad
	J^{\rm scalar}_\text{abs} = \frac{1}{\pi} \int_0^\infty d\omega\,\omega\sum_{\ell m} m \left| W^H_{\ell m,0}\right|^2\,. \label{eq:absEJscl}
\end{align}
The relative factor $2$ comes from the fact that we have two propagating degrees of freedom for the vector and a single one for the scalar field.  Again the formulae for the radiated quantities are identical to these up to replacing $W^H_{\ell m,+1}$ with $W^\infty_{\ell m,-1}$ and $W^H_{\ell m,0}$ with $W^\infty_{\ell m,0}$.

\section{ Absorbed fluxes at leading PM order } \label{sec:computation_all}

In this section, we derive explicit expressions for the waveforms and fluxes at the horizon to leading PM order.

\subsection{The PM limit} \label{sec:LO_PM}

In the limit $x \to 0$ for generic $y$, the solutions $\mathfrak{R}_\text{in}$,  $\mathfrak{R}_\text{up}$ given in (\ref{eq:frakRinfrakRupGENERAL}) drastically simplify. Indeed, in this limit, the dictionary \eqref{eq:dictionary} becomes\begin{equation}
	m_1 \to 0\,,\quad
	m_{2,3} \to s\,,\quad
	a \to \ell+\tfrac{1}{2}
\end{equation}
 and ${\cal F}_{\rm inst},\, {{g}_+/{g}_-} \to 0$. The limit $x\to0$ should be performed after setting $s=0,\pm1, \pm2$ but keeping $\ell$ generic and only letting $\ell$ approach integer values at the very end.
 The leading-order expressions can be summarized as follows,
\begin{equation}\label{eq:Rinleadingx}
	\mathfrak{R}_\text{in}
	\underset{x\to0}{\sim}
	-
	\frac{\Gamma(\ell+s+1)}{\Gamma(2\ell+2)}
	\,e^{-\frac{y}{2}} (-y)^{\ell-s}\,{}_1F_1(\ell+1-s,2\ell+2;y)
\end{equation}
and\footnote{The argument of the second gamma function in the external denominator still contains $x$ because, when $s = 1,2$, we have to first set $s$ to one of these values and then perform the $x \to 0$ limit.}
\begin{equation}\label{eq:Rupleadingx}
	\begin{split}
		\mathfrak{R}_\text{up}
		&\underset{x\to0}{\sim}
		-\frac{ x^{\ell+1+s} \Gamma(\ell+1) \Gamma(\ell+1-s)}{ \Gamma(2\ell+2)\Gamma(1-s-x)}\,
		\,e^{-\frac{y}{2}}
		y^{-s-1-\ell}
		\Big[
		{}_1F_1(-s-\ell,-2\ell;y)
		\\
		&+
		\frac{y^{2\ell+1}\Gamma(-2\ell)\Gamma(\ell+1+s)}{\Gamma(s-\ell)\Gamma(2\ell+2)}
		{}_1F_1(\ell+1-s,2\ell+2;y)
		\Big].
	\end{split}
\end{equation}
These leading-order solutions can be also obtained by directly solving the Teukolsky equation (\ref{eq:TeukolskyHom}) in the limit $M\omega \ll 1$. 
Seen from the exterior region,  the horizon collapses to the origin in this limit, leading to a smooth Minkowski space, and consequently the incoming boundary conditions are replaced by the condition that $\mathfrak{R}_\text{in}$ be regular at the origin
$r=0$. This leads to~\eqref{eq:Rinleadingx}. The normalization factor
 $C_{\rm in}$ is determined by imposing the correct asymptotics as $r\to \infty$.  For $\mathfrak{R}_\text{up}$, the solution ~\eqref{eq:Rupleadingx} is determined by imposing upgoing boundary conditions at infinity, up to an overall 
 normalization $C_{\rm up}$. To determine $C_{\rm up}$ one needs to solve  \eqref{eq:TeukolskyHom} also in the interior region, $M,r\ll \omega^{-1}$, and match its asymptotics against the exterior solution.

Note that, while the ingoing solution $\mathfrak{R}_\text{in}$ starts at order $x^0$, hence $G^0$, for any $\ell$, the upgoing one $\mathfrak{R}_\text{up}$ goes like $G^{\ell+1+s}$ for $s\le0$ and like $G^{\ell+2+s}$ for $s=1,2$. Consequently, low multipoles with $\ell=|s|$ dominate in the PM expansion.
 The leading contributions are therefore,
 for $|s|=2$,
\begin{subequations}
	\begin{align}
		\mathfrak{R}^{\ell=-s=2}_\text{up} 
		&\underset{x\to0}{\sim}
		-\frac{x e^{y/2} \left(y^4-4 y^3+12 y^2-24 y+24\right)}{120 y}\,,
		\\
		\mathfrak{R}^{\ell=s=2}_\text{up} 
		&\underset{x\to0}{\sim}
		-\frac{x^6 e^{y/2}}{60 y^5}\,,
	\end{align}
\end{subequations}
 for $|s|=1$,
\begin{subequations}
	\begin{align}
		\mathfrak{R}^{\ell=-s=1}_\text{up} 
		&\underset{x\to0}{\sim}
		-\frac{x e^{y/2} \left(y^2-2 y+2\right)}{6 y}\,,
		\\
		\mathfrak{R}^{\ell=s=1}_\text{up} 
		&\underset{x\to0}{\sim}
		\frac{x^4 e^{y/2}}{6 y^3}\,,
	\end{align}
\end{subequations}
 and for $s=0$
\begin{subequations}
	\begin{align}
		\mathfrak{R}^{\ell=s=0}_\text{up} 
		&\underset{x\to0}{\sim}
		-\frac{x e^{y/2}}{y}\,.
		 	\end{align}
\end{subequations}
 In the last case, we will also need the solution for $\ell=1$, 
\begin{subequations}
	\begin{align}
		\mathfrak{R}^{\ell=1,s=0}_\text{up} 
		&\underset{x\to0}{\sim}
		\frac{x^2 e^{y/2} (y-2)}{12 y^2}\,,
	\end{align}
\end{subequations}
 which, as we will see, gives the leading contribution to the absorbed angular momentum. 

\subsection{Integral over the geodesics}
\label{sec:geodesic}
 
In the computation of $Z_{\ell m,s}^H$, we have to perform the integral along the particle's trajectory, which we now discuss.
As already said, we consider perturbations generated by a particle of mass $\mu$ moving along a geodesic on the equatorial plane of a Schwarzschild geometry. We denote by 
\begin{equation}
E=\mu \, f(r)\,\frac{dt}{d\tau} = \mu\,\sigma = \mu \sqrt{1+p_\infty^2}  \,, \qquad 
L=\mu r^2\, \frac{d\phi}{d\tau}  = \mu \, j\,   \label{sigmapinf}
\end{equation}
the energy and angular momentum of the particle and by
\begin{equation}
b = \frac{j}{p_\infty}
\end{equation} 
 its impact parameter. 
In terms of these variables, the geodesic equations can be  written as
\begin{equation}\label{eq:drdlambda}
	\pm \frac{dr}{\sqrt{p_\infty^2-2\mathcal{V}(r)}} =  d\tau\,,
	\qquad
	\pm \frac{j\,dr}{r^2\sqrt{p_\infty^2-2\mathcal{V}(r)}} = d\phi\,.
\end{equation}
where
\begin{equation}\label{eq:Veff}
	\mathcal{V}(r) = \frac{j^2}{2r^2} - \frac{M}{r} - \frac{M j^2}{r^3}\,.
\end{equation}
is the effective potential. We focus on hyperbolic-like encounters, so we take $p_\infty$ real and  $\sigma > 1$. 
 The geodesic equations \eqref{eq:drdlambda} can be easily solved perturbatively  as an expansion in  $M/b$. 
  To leading order one finds the free (straight-line) trajectory  
\begin{equation}\label{eq:vecFree0}
	t(\tau) = \tau \sqrt{1+p_\infty^2} +\cdots \,,\quad
	r(\tau) =  \sqrt{b^2+ p_\infty^2 \tau^2} +\cdots \,,\quad
	\phi(\tau) = \arctan \frac{\tau p_\infty }{b}+\cdots 
\end{equation}
 with dots denoting terms suppressed by $M/b$ that will be omitted in the following because these corrections are suppressed by higher powers of $G$.
 Introducing the variable $\kappa$ such that
\begin{equation}
t = \frac{b\,  \kappa}{p_\infty}
\end{equation}
one finds
\begin{equation}\label{eq:vecFree}
	r(\kappa) = b \sqrt{1+  \ft{\kappa^2}{1+p_\infty^2} }\,,\qquad
	e^{{\rm i} \phi(\kappa)}  = { \sqrt{1+p_\infty^2}+{\rm i} \kappa\over \sqrt{1+\kappa^2+p_\infty^2} } \, .
\end{equation}
We plug this into  (\ref{eq:ZbL}), expand the integrand in the limit $p_\infty \to 0$ and then evaluate the integral order by order in $p_\infty$. As we are going to see below, we can then resum the $p_\infty$ expansion, furnishing at the end results that are valid for generic $p_\infty$. By doing this, we find
 \begin{equation}
	Z^H_{\ell m,s} 
	=
	\frac{b}{p_\infty} \int_{\mathbb{R}}  d\kappa \,e^{i \mathsf{u} \kappa - i m \phi(\kappa)}
	\sum_{i=0}^2 b^i_{\ell m}\,\mathcal{L}_{i} [\mathfrak{R}_{\text{up}}( r (\kappa) )] =\sum_{n}  \int_{\mathbb{R}}  d\kappa \,e^{i \mathsf{u} \kappa} \frac{P_{\ell m n}(\kappa,\mathsf{u})}{(\kappa^2+1)^{\nu_{\ell m n}} }    p_\infty^n
\end{equation}
with 
\be
\mathsf{u} = \frac{\omega b}{p_\infty} \, ,
\ee
and where $P_{\ell m n}(\kappa,\mathsf{u})$ are some polynomials in $\kappa$ and $\mathsf{u}$. We notice that monomials in $\kappa$ in each polynomial can be replaced by derivatives $\kappa \to -i \partial_{\mathsf{u}}$ upon integration by parts.
 For this reason, the remaining integral takes the master  form
\begin{equation}
 \int_{\mathbb{R}} \frac{e^{i \mathsf{u} \kappa}}{(\kappa^2+1)^{\nu}} \,  d \kappa = \frac{2^{\frac{3}{2}-\nu} \sqrt{\pi}\, \mathsf{u}^{\nu-\frac{1}{2}}}{\Gamma(\nu)} \, K_{\nu-\frac{1}{2}}(\mathsf{u})\, .
\end{equation}

\subsection{Gravitational case}
In this section, we compute the waveform, the absorbed energy and angular momentum for the gravitational scenario. We start from the multipolar coefficients $W^H_{\ell m,+2}$ derived from $Z_{\ell m,-2}^H$ via \eqref{eq:WHZa}. The same result is obtained also from ${Z}_{\ell m,+2}^H$ via \eqref{eq:WHtildeZ}.\\
The leading PM contributions for $s=-2$ come from the lowest multipole $\ell=2$. For $m=2$ let us explicitly see what happens.
Performing the integral over the trajectories, at each order in $p_\infty$, we find the following result 
\begin{align}
Z_{22,-2}^{H} (\mathsf{u}) &=  G M \mu \sqrt{\frac{\pi}{5}}  \left[  \left(-\frac{4 \mathsf{u}^2}{b^2 p_\infty} - \frac{2 p_\infty \mathsf{u}^2}{b^2} \right) K_0 (\mathsf{u}) \right. \nn\\
& \left.  +  \left(-\frac{2 \mathsf{u}(1+2\mathsf{u})}{b^2 p_\infty} - \frac{2 p_\infty \mathsf{u}(2+\mathsf{u})}{b^2} + \frac{p_\infty^3 \mathsf{u}^2}{2 b^2} + \cdots \right) K_1 (\mathsf{u}) \right]  .
\end{align}
We notice that the coefficient of $K_0(\mathsf{u})$ does not receive higher order corrections in $p_\infty$. Resumming the velocity expansion for the coefficient of $K_1(\mathsf{u})$, we arrive at
\begin{equation} \label{eq:resultWh22m2}
Z_{22,-2}^{H}(\mathsf{u}) = G M \mu \sqrt{\frac{\pi}{5}}\left(-\frac{2 \mathsf{u}^2 (p_\infty^2+2)}{b^2 p_\infty} K_0(\mathsf{u})-\frac{2 \mathsf{u}(1 + 2 \mathsf{u} \sqrt{1+p_\infty^2} + 2 p_\infty^2)}{b^2 p_\infty} K_{1}(\mathsf{u})\right).
\end{equation}
We checked the resummed expressions explicitly up to relative $\mathcal{O}(p_\infty^{20})$. We observe that only the first line of \eqref{eq:Rupleadingx} gives nonzero contributions, while those from the second line vanish after the $\kappa$-integration at each order in $p_\infty$. The same pattern continues also for the other $Z_{\ell m, s\le0}^H$ presented below.

By repeating the same procedure for all the other modes and expressing all these quantities in terms of $\sigma$, via \eqref{sigmapinf}, we obtain
\begin{subequations}\label{eq:resultWN}
	\begin{align}
	        Z_{2(\pm 2),-2}^{H}(\mathsf{u}) 
	        &= 
	        G M \mu \sqrt{\frac{\pi}{5}}
	        \left(
	        -\frac{2 \mathsf{u}^2 (\sigma^2+1)}{b^2 \sqrt{\sigma^2-1}} K_0(\mathsf{u})
	        -
	        \frac{2 \mathsf{u}(-1 \pm 2 \mathsf{u} \sigma + 2 \sigma^2)}{b^2 \sqrt{\sigma^2-1}} K_{1}(\mathsf{u})\right)\, ,
	        \\
		Z_{2 (\pm 1),-2}^{H}(\mathsf{u}) 
		&= 
		G M \mu \sqrt{\frac{\pi}{5}}
		\left(
		\mp
		\frac{4 i \mathsf{u}^2 \sigma}{b^2}\,K_0(\mathsf{u})
		-
		\frac{4 i \mathsf{u} (\mathsf{u} \pm 2 \sigma )}{b^2}
		\, K_1(\mathsf{u})
		\right),
		\\
		Z_{2 0,-2}^{H}(\mathsf{u})
		&= 
		G M \mu \sqrt{\frac{24\pi}{5}}
		\left(
		\frac{\mathsf{u}^2 \sqrt{\sigma^2-1}}{b^2}K_0(\mathsf{u}) 
		+
		\frac{\mathsf{u} (2 \sigma^2-1)}{b^2 \sqrt{\sigma^2-1}}
		K_1(\mathsf{u})
		\right),
	\end{align}
\end{subequations}
Using (\ref{eq:WHZa}), that at leading PM order reduces to
\begin{subequations}
	\begin{align}  \label{eq:relation_W_Z_LO}
		W_{\ell m,+2}^{H}
		&=
		-
		\frac{16 M^2}{ G \ell (\ell^2-1)(\ell+2)}   
		Z_{\ell m,-2}^{H} \,,
	\end{align}
\end{subequations}
we thus obtain the waveform coefficients. We note that these are exponentially suppressed for large $\mathsf{u}$, i.e.~$\omega \gg b^{-1}$, while they go to a constant for small $u$, i.e.~in the soft limit $\omega \ll b^{-1}$. Contrary to the usual waveforms at infinity, they do not exhibit a $1/\omega$ pole in the soft regime.\\
Inserting the waveform coefficients \eqref{eq:relation_W_Z_LO} into \eqref{eq:absEJgr}, we find the following expressions for the amount of absorbed energy and angular momentum per unit of frequency
\begin{align}
\frac{d E_{\mathrm{abs}}}{d \mathsf{u}} & = \frac{G}{\pi^2} \frac{\sqrt{\sigma^2-1}}{b} \left(\frac{\mathsf{u} \sqrt{\sigma^2-1}}{b}\right)^2 \sum_{m=-2}^2 |W_{2m,+2}^H|^2 =  \notag \\ 
& = \frac{128 G^7 \mu^2 M_{\text{BH}}^6 \mathsf{u}^4 \sqrt{\sigma^2-1}}{45 \pi b^7} \left[\mathsf{u}^2 \left(1-2\sigma^2+2 \sigma^4\right) K_0(\mathsf{u})^2 + \right. \notag \\
& \left. + \mathsf{u} \left(1-8 \sigma^2+8 \sigma^4\right) K_0(\mathsf{u}) K_1(\mathsf{u}) + (1-8 \sigma^2 + 8 \sigma^4 + \mathsf{u}^2(2 \sigma^2-1)) K_1(\mathsf{u})^2 \right]\, , \\
\frac{d J_{\mathrm{abs}}}{d \mathsf{u}} & = \frac{G}{\pi^2} \frac{\sqrt{\sigma^2-1}}{b} \left(\frac{\mathsf{u} \sqrt{\sigma^2-1}}{b}\right) \sum_{m=-2}^2 m |W_{2m,+2}^H|^2 = \notag \\
& = \frac{256 G^7 \mu^2 M_{\text{BH}}^6 \mathsf{u}^4 \sigma K_1(\mathsf{u}) \left[2 \mathsf{u} \sigma^2 K_0(\mathsf{u}) + \left(4 \sigma^2-3\right) K_1(\mathsf{u})\right]}{45 \pi b^6}\, .
\end{align}
We note that, due to the absence of a $1/\omega$ pole in $W_{\ell m,+2}^H$, the absorbed angular momentum is insensitive to static $\delta(\omega)$ contributions, unlike the one emitted to infinity \cite{Manohar:2022dea,DiVecchia:2022owy,Heissenberg:2024umh,Heissenberg:2025ocy}. This is also supported by the explicit check in Section~\ref{sec:nonrel_check} below.\\
Finally, performing the  integrals we obtain the (total) absorbed energy and angular momentum
\begin{subequations}
\begin{align}\label{eq:Eabs}
	E_\text{abs} &= \frac{5 \pi  G^7 \mu ^2 M_\text{BH}^6 \sqrt{\sigma ^2-1} \left(21 \sigma ^4-14 \sigma ^2+1\right)}{16 b^7}\, , \\
	J_\text{abs}& = \frac{\pi  G^7 \mu ^2 M_\text{BH}^6 \sigma  \left(7 \sigma ^2-3\right)}{2 b^6}\,.\label{eq:Jabs}
\end{align}
\end{subequations}
The result for the energy agrees with~\cite{Goldberger:2020wbx,Jones:2023ugm}, while the one for the absorbed angular momentum is new. The total absorbed energy is one of the dissipative observables that have recently been analysed in~\cite{Warburton:2025ymy} with a numerical self-force approach to gravitational scattering. At large values of the impact parameter, the results of~\cite{Warburton:2025ymy} are consistent with the analytic formula above for the absorbed energy, see Fig.~6 of~\cite{Warburton:2025ymy}. The author performed\footnote{We would like to thank Niels Warburton for sharing this result, see also~\cite{Warburton_nordita}.} a similar numerical check for the absorbed angular momentum finding, at leading PM order, agreement with~\eqref{eq:Jabs}.

The results \eqref{eq:resultWh22m2}--\eqref{eq:Jabs} are valid for relativistic velocities of the light object, i.e.~for generic $\sigma>1$. 
Moreover, although we have derived them by applying leading-order perturbation theory in the probe limit, we expect them to be exact in the mass ratio $\mu/M_\text{BH}$ to leading order in the PM expansion, due to the fact that PM expressions are simple polynomials in the masses and leading PM ones are actually  \emph{monomials} \cite{Damour:2019lcq,Heissenberg:2025fcr}. Indeed, \cite{Goldberger:2020wbx,Jones:2023ugm} derived the absorbed energy for generic masses, to leading order in $G$, and verified that \eqref{eq:Eabs} provides the complete result at that order. 

\subsection{Vector case}

In the vector case $s=-1$, the leading contribution comes from $\ell=1$. Proceeding as before
 one finds
 \begin{equation}
	Z^{H}_{1(\pm1),-1} = 2 \mathsf{u} q_e \, M \sqrt{\frac{2\pi}{3}} \frac{K_0(\mathsf{u})\pm \sigma K_1(\mathsf{u})}{b \sqrt{\sigma^2-1}}\,,
	\qquad
	Z^{H}_{10,-1} = 4 i  M  q_e \sqrt{\frac{\pi}{3}}\frac{ \mathsf{u} \, K_1(\mathsf{u})}{b}
\end{equation}
 Using (\ref{eq:WHZb}), at leading PM order
\begin{equation}
	W^{H}_{\ell m,+1}(\omega) = \frac{4M}{\ell(\ell+1)}\,Z^{H}_{\ell m,-1}(\omega)\, ,
\end{equation}
 one finds that the rate of absorbed energy and angular momentum per unit of frequency are, from \eqref{eq:absEJem}
\begin{align}
\frac{d E^{\rm vector}_{\mathrm{abs}}}{d \mathsf{u}} & = \frac{2}{\pi} \frac{\sqrt{\sigma^2-1}}{b} \left(\frac{\sqrt{\sigma^2-1} \mathsf{u}}{b}\right)^2 \sum_{m=-1}^1 |W_{1m,+1}^H|^2 \notag \\ 
& = \frac{128 G^4 q_e^2 M_{\text{BH}}^4 \mathsf{u}^4 \sqrt{\sigma^2-1}}{3 b^5} \left[K_0(\mathsf{u})^2 + \left(2 \sigma^2-1\right) K_1(\mathsf{u})^2\right]\, , \\
\frac{d J^{\rm vector}_{\mathrm{abs}}}{d \mathsf{u}} & = \frac{2}{\pi} \frac{\sqrt{\sigma^2-1}}{b} \left(\frac{\sqrt{\sigma^2-1} \mathsf{u}}{b}\right) \sum_{m=-1}^1 m |W_{1m,+1}^H|^2  \notag
\\
&= \frac{256 G^4 q_e^2 M_{\text{BH}}^4  \mathsf{u}^3 \sigma}{3 b^4} K_0(\mathsf{u}) K_1(\mathsf{u})\,.
\end{align}
Integrating, we finally obtain
\begin{equation}
	E^{\rm vector}_\text{abs} = \frac{3 G^4 M_\text{BH}^4 \pi^2 q_e^2 \sqrt{\sigma^2-1} \left(5 \sigma ^2-1\right)}{2 b^5}\,,
	\qquad
	J^{\rm vector}_\text{abs} = \frac{4 G^4 M_{\text{BH}}^4 \pi^2 q_e^2\sigma}{b^4}\,.
\end{equation}
This result for $E^{\rm vector}_\text{abs}$ agrees with (3.38) of \cite{Jones:2023ugm}, after rescaling the coupling according to\footnote{We explicitly checked that this dictionary between the two sets of conventions also works for other observables, such as the electromagnetic deflection angle. Similarly it happens for the scalar case mentioned below.\label{footnote:matching}}
\begin{equation}
Q_e = 4 \pi q_e
\end{equation}
to match conventions ($Q_e$ is the coupling of that reference), while $J^{\rm vector}_\text{abs}$ is new.\\

\subsection{Scalar case}
 
 In the scalar case, the leading PM contribution to the absorbed energy comes from the
  $\ell=0$ mode. One finds
\begin{equation}
	W^{H}_{00,0} (\mathsf{u})= 8 \sqrt{\pi} G M_\text{BH} q \, \frac{K_0(\mathsf{u})}{\sqrt{\sigma^2-1}}\,,
\end{equation}
from which we obtain the leading-order absorbed energy per unit of frequency and the total one from \eqref{eq:absEJscl}
\begin{align}
\frac{d E^{\rm scalar}_{\text{abs}}}{d \mathsf{u}} & = \frac{1}{\pi} \frac{\sqrt{\sigma^2-1}}{b} \left(\frac{\sqrt{\sigma^2-1} \, \mathsf{u}}{b}\right)^2 |W_{00,0}^H|^2 = \frac{64 G^2 M_{\text{BH}}^2 q^2  \mathsf{u}^2 \sqrt{\sigma^2-1} K_0(\mathsf{u})^2}{b^3}\, , \\
E^{\rm scalar}_\text{abs} &= \frac{2 G^2 M_\text{BH}^2  q^2 \pi^2 \sqrt{\sigma ^2-1}}{b^3}\,. \label{eq:Eabs0}
\end{align}
Instead, the $\ell=0$ mode does not contribute to the absorbed angular momentum and the leading-order contribution comes from the $\ell=1$ modes, 
\begin{equation}
	W^{H}_{1(\pm1),0}(\mathsf{u}) = -4 G^2 M^2_\text{BH} q \sqrt{\frac{2\pi}{3}} \frac{\mathsf{u}\left(\sigma K_0(\mathsf{u})\pm  K_1(\mathsf{u})\right)}{b\sqrt{\sigma^2-1}}\,,
	\qquad
	W^{H}_{10,0}(\mathsf{u}) = 0\,.
\end{equation}
Then from~\eqref{eq:absEJscl} we obtain, at leading order in $G$,
\begin{align}
\frac{d J^{\rm scalar}_{\text{abs}}}{d \mathsf{u}} &= \frac{1}{\pi} \frac{\sqrt{\sigma^2-1}}{b} \left(\frac{\sqrt{\sigma^2-1} \, \mathsf{u}}{b}\right) \sum_{m=-1}^1 m |W_{1m,0}^H|^2 = \frac{128 G^4 M_{\text{BH}}^4 q^2 \mathsf{u}^3 \sigma K_0(\mathsf{u}) K_1(\mathsf{u})}{3 b^4}\, , \\
J^{\rm scalar}_\text{abs} & =\frac{2 G^4 M^4 \pi^2 q^2 \sigma}{b^4}\,.
\end{align}
The result for $E^{\rm scalar}_\text{abs}$ in \eqref{eq:Eabs0} agrees with  (3.36) of \cite{Jones:2023ugm}, after rescaling the coupling according to (see Footnote~\ref{footnote:matching})
\begin{equation}
Q_s = 8 \pi \mu \, q
\end{equation}
to match conventions ($Q_s$ is the coupling of that reference). Instead the result for $J^{\rm scalar}_\text{abs}$ is new and is further suppressed by an extra factor of $(GM_\text{BH}/b)^2$ with respect to the naive expectation since it follows from the $\ell=1$ horizon waveforms.\\

\subsection{Checks of the gravitational results in a nonrelativistic limit} \label{sec:nonrel_check}
We can obtain cross-checks for both results \eqref{eq:Eabs} and \eqref{eq:Jabs} in the nonrelativistic or PN limit by using the horizon radiation-reaction force obtained in Section~3.2 of \cite{Goldberger:2020wbx} to leading order in the velocity, which we employ in the probe limit $\mu \ll M_\text{BH}$,\footnote{To match notation with \cite{Goldberger:2020wbx}, $m_1|_{\rm there}=M_\text{BH}$, $m_2|_{\rm there}=\mu$, $\vec{x}_1=0$, $\vec{x}_2=\vec{r}$, so $\vec{x}=\vec{x}_1-\vec{x}_2=-\vec{r}$ (so $\vec{v}\,|_\text{there}=-\vec{v}\,|_\text{here}$) and we focus on the force exerted on the probe, $\vec{F}_2 = \vec{F}$. This is because the heavy object does not move and sits in the origin, so the force exerted by the probe on it produces no work and no torque.}
\begin{equation}\label{eq:HRRforce}
	\vec{F} = - \frac{32}{5}\,\frac{G^7M_\text{BH}^4\mu^2}{{r}^8}
	\left(
	\vec{v}+\frac{2\vec{v}\cdot \vec{r}}{{r}^2}\,\vec{r}
	\right).
\end{equation}
Here, $\vec{v} = \frac{d}{dt}\,\vec{r}$ and to leading order in $G$ and $p_\infty$ we can work with the free trajectory \eqref{eq:vecFree} in the small-$p_\infty$ approximation, i.e. $\vec{r} = (x,y,z)$,
\begin{equation}
	x = r(\kappa) \cos\phi(\kappa) = b + \mathcal{O}(p_\infty^2)\,,\qquad
	y = r(\kappa) \sin\phi(\kappa) =  p_\infty t  + \mathcal{O}(p_\infty^3)\,,\qquad
	z = 0\,.
\end{equation} 
The work exerted by the dissipative force is thus 
\begin{equation}
	\Delta E = \int \vec{v}\cdot \vec{F}\,dt\,.
\end{equation}
The $t$-integral is elementary and yields
\begin{equation}
	\Delta E = -\frac{5 \pi  G^7 \mu ^2 M_\text{BH}^6}{2 b^7}\,p_\infty 
\end{equation}
and this agrees with the small-$p_\infty$ expansion of \eqref{eq:Eabs}, since $\Delta E = -E_\text{abs}$. Similarly, we can integrate the torque exerted by the force to obtain the change in angular momentum,
\begin{equation}
	\Delta \vec{J} = \int \vec{r} \times \vec{F}\,dt\,,
\end{equation}
which yields $\Delta \vec{J} = (\Delta J)\,\hat{z}$ with
\begin{equation}
	\Delta J = -\frac{2 \pi  G^7 \mu ^2 M_\text{BH}^6}{b^6}\,.
\end{equation}
This agrees with the PN expansion of the PM result \eqref{eq:Jabs}, owing to $\Delta J = - J_\text{abs}$.

\section{Conclusions} \label{sec:conclusions}

In this paper we computed, using black hole perturbation theory,  the flux of energy and angular momentum absorbed by the horizon of a Schwarzschild black hole during scattering at the leading order in the PM expansion. To do so, we use an ``exact'' PM formula, recently derived in  \cite{Cipriani:2025ikx},  describing the solution of confluent Heun equation, at leading order in the PM expansion, as a hypergeometric function with argument $2{\rm i} \omega r$. 
Our results for scalar, vector and gravitational energy absorption by the black hole horizon are finally compared with those obtained for the increase of the black hole mass obtained in \cite{Goldberger:2020wbx,Jones:2023ugm} and we find perfect agreement. 

The results for the angular momentum absorption at leading PM order are new and, in the  nonrelativistic limit, the gravitational one matches the loss of angular momentum obtained using the dissipative force of~\cite{Goldberger:2020wbx}. It would be interesting to further study this observable in the amplitude-EFT beyond the small-velocity limit, which may be possible using the approach of~\cite{Aoude:2023fdm,Aoude:2024jxd,Gatica:2025uhx}. Once again, the variation of the black hole spin ought to match our result~\eqref{eq:Jabs} for the absorbed angular momentum, turning a Schwarzschild black hole in the initial state into a Kerr black hole in the final state. The techniques developed in this paper should extend straightforwardly to the evaluation of higher order terms in the PM expansion of the absorbed quantities. While these corrections are highly suppressed, they can provide a check for possible extensions of the EFT analysis.

Since our analysis does not rely on a PN approximation, one may be tempted to take the ultrarelativistic limit $\sigma\to\infty$. However, this limit is known to be subtle. First of all, some care is needed when dealing with the order of the probe and the high velocity limits. By taking the probe limit first, let us restrict to a regime where the ultrarelativistic light particle is still much less energetic than the black hole, 
 $\mu \sigma\ll M_\text{BH}$. Then, from~\eqref{eq:Eabs}, we see that, in the gravitational case, the fraction of the probe's energy that is absorbed by the black hole, $E_\text{abs}/(\mu \sigma)$, scales like $ \mu\, \Theta^7 \sigma^4/M_\text{BH}$. Here $\Theta$ is the deflection angle of the massless particle in the rest frame of the black hole, $\Theta\sim\mathcal{O}(GM_\text{BH}/b)$, to be kept fixed and small. Clearly the fraction $E_\text{abs}/(\mu \sigma)$ ought to be smaller than one, while the above estimate grows like $\sigma^4$. This means that our result is indeed reliable for parametrically large $\sigma$, but only  below the bound  $\sigma \lesssim \mathcal{O} (\mu \Theta^7/M_\text{BH})^{-1/4}$. It is clear that, beyond this threshold, the results are no longer sensible and moreover we cannot assume that~\eqref{sigmapinf} represent conserved quantities, so that one needs a more precise analysis. 
  
Since, by the EFT analysis \cite{Goldberger:2020wbx,Jones:2023ugm}, we know that~\eqref{eq:Eabs} is in fact exact in the masses, we can also relax the constraints on the relative size of $\mu$ and $M_\text{BH}$. We can then take $\sigma\to\infty$ while keeping the center-of-mass deflection angle, $\Theta_\text{CM}$, fixed and small. Doing so, we find that the fraction of center-of-mass energy absorbed by the black hole is $E_\text{abs}/E_\text{CM} \sim \mathcal{O}(\Theta_\text{CM}^7 \sigma)$, where now $\Theta_\text{CM}\sim  \mathcal{O}(GE_\text{CM}/b)$  with $E_\text{CM}\sim  \mathcal{O}(\sqrt{M_\text{BH}\mu \sigma})$. In this regime, we thus find the bound $\sigma \lesssim   \mathcal{O}(\Theta_\text{CM}^{-7})$, beyond which a more precise analysis is required.

The origin of the bad ultrarelativistic behavior of our results can be traced back to the high-requency regime, which in principle contributes to the integrated absorbed quantities, see~\eqref{eq:absEJgr}. In practice frequencies larger than $p_\infty/b$ are irrelevant as the Bessel functions in~\eqref{eq:resultWN} make the integrands exponentially suppressed. However, when $\sigma \gtrsim 1/\Theta$, this suppression kicks in for frequencies larger than $1/M$ which are not reliably described in our PM approximation $|x|=4M\omega \ll 1$.
 
 The absorbed angular momentum \eqref{eq:Jabs} shares a pathological behavior in the first regime, where $J_\text{abs}/(\mu b \sigma)$ scales like $\mu\, \Theta^7\sigma^2/M_\text{BH}$, but is intriguingly well behaved in the second one, $J_\text{abs}/(p_\text{CM}b) \sim \mathcal{O}(\Theta_\text{CM}^7 \sigma^{-1})$.
To summarize, while our results are indeed reliable for parametrically large velocities, it is likely that the naive PM approximation (i.e.~expanding in $G$ first and then taking the large $\sigma$ limit) eventually breaks down and one needs to partially resum it in order to access a reliable result in the strict ultrarelativistic regime \cite{Kovacs:1978eu,Goldberger:2020wbx,DiVecchia:2022nna,Alessio:2024onn}. See~\cite{Rothstein:2024nlq,Alessio:2025isu,Alessio:2026bdi} for recent progress in the EFT for this regime.

Another conceptually important question that would be worth analysing is the role of horizon supertranslations  \cite{Donnay:2015abr,Donnay:2016ejv,Mao:2023dsy}. Let us also note that the $W_{\ell m,+2}^H$ in principle transforms nontrivially under these supertranslations and it will be interesting to further explore the connection between absorbed energy and angular momentum and such asymptotic symmetries.

Going beyond the case of the Schwarzschild geometry presented here, there are various extensions that it is possible to study by using the approach of this paper such as the cases of Kerr (a PM formula for the absorption has recently derived with the EFT method in \cite{Bautista:2024emt}) and of higher dimensional black holes~\cite{Bianchi:2021xpr,Bianchi:2021mft,Akhtar:2025nmt}. It would be interesting also to go beyond the realm of black holes and consider topological stars \cite{Bah:2021jno,Heidmann:2022ehn,Bianchi:2023sfs,Cipriani:2024ygw,Bianchi:2025uis,Bah:2025vbr,Heidmann:2025pbb,Bena:2025usu}.  Gravitational waves on topological star geometry have been recently studied in the PN approximation in \cite{Bianchi:2024vmi,Bianchi:2024rod,Bianchi:2025aei,Bianchi:2025ydq}. In a more formal context, the Heun equation appears also in the study of asymptotically Anti de Sitter solutions that describe precisely identified heavy states on the dual conformal field theory side~\cite{Aprile:2025hlt}. Extending our approach to such cases may highlight some yet unexplored feature of the dynamics of heavy state in the AdS/CFT duality.

\subsection*{Acknowledgements}
It is a pleasure to thank Massimo Bianchi, Stefano De Angelis, Giorgio Di Russo, Davide Fioravanti, Stefano Foffa, Cristoforo Iossa, Ira Rothstein, Riccardo Sturani, Davide Usseglio, Juan Valiente-Kroon and Niels Warburton for enlightening discussions. AC would like to thank IPhT and IHES for the very kind hospitality during various stages of this work. RR is partially supported by the UK EPSRC grant ``CFT and Gravity: Heavy States and Black Holes" EP/W019663/1 and by the Science and Technology Facilities Council (STFC) Consolidated Grant ST/X00063X/1 ``Amplitudes, Strings \& Duality''.
  
\appendix

\section{Conventions and background geometry}
\label{sec:app_A}

We work with the mostly-plus signature. 
We define the Levi--Civita connection coefficients by
\begin{equation}
	\Gamma^{\mu}_{\alpha\beta} = \frac{1}{2} g^{\mu\nu}\left(
	\partial_\alpha g_{\nu\beta} + \partial_\beta g_{\nu\alpha} - \partial_\nu g_{\alpha\beta}
	\right),
\end{equation}
the Riemann curvature tensor by 
\begin{equation}
	R\indices{^\mu_{\nu\alpha\beta}} = \partial^{\phantom{\mu}}_\alpha \Gamma^{\mu}_{\nu\beta} - \partial^{\phantom{\mu}}_\beta \Gamma^{\mu}_{\nu\alpha} + \Gamma^\mu_{\rho\alpha}  \Gamma^\rho_{\beta \nu} - \Gamma^\mu_{\rho\beta}  \Gamma^\rho_{\alpha \nu}
\end{equation}
and the Ricci tensor and curvature scalar by
\begin{equation}
	R_{\mu\nu}  = R\indices{^\alpha_{\mu\alpha\nu}}\,,\qquad
	R = g^{\mu\nu} R_{\mu\nu}\,.
\end{equation}
The Weyl tensor is given by
\begin{equation}
	\begin{split}
		C_{\alpha\beta\rho\sigma}
		&=
		R_{\alpha\beta\rho\sigma}
		+\frac{1}{D-2}(
		R_{\alpha \sigma} g_{\beta\rho}
		-
		R_{\alpha \rho} g_{\beta\sigma}
		+
		R_{\beta \rho} g_{\alpha \sigma}
		-
		R_{\beta\sigma} g_{\alpha \rho})
		\\
		&+
		\frac{1}{(D-1)(D-2)}
		\,R
		(
		g_{\alpha \rho} g_{\beta \sigma}
		-
		g_{\alpha \sigma} g_{\beta\rho}
		)
	\end{split}
\end{equation}
and, once a tetrad $\ell$, $n$, $m$, $\bar{m}$ is fixed, we define the two Newman--Penrose scalars that are relevant here as follows 
\begin{equation}\label{eq:defPsi4Psi0}
	\Psi_0 = C_{\mu\nu\rho\sigma} \ell^\mu {m}^\nu \ell^\rho {m}^\sigma\,,
	\qquad
	\Psi_4 = C_{\mu\nu\rho\sigma} n^\mu \bar{m}^\nu n^\rho \bar{m}^\sigma\,.
\end{equation}

The Schwarzschild metric \eqref{eq:Schwarzschild} is a vacuum solution of the Einstein equations,
\begin{equation}
	R_{\mu\nu} = 0\,,
\end{equation}
(with a  singularity at $r=0$).
We denote by
\begin{equation}\label{eq:roundspheremetric}
	\gamma_{AB} = \left(\begin{matrix}
		1 & 0 \\ 0 & (\sin\theta)^2
	\end{matrix}\right),
	\qquad
	\gamma^{AB} = \left(\begin{matrix}
		1 & 0 \\ 0 & (\sin\theta)^{-2}
	\end{matrix}\right)
\end{equation}
the metric on the round sphere in $\theta,\phi$ coordinates and its inverse.
It can be convenient to introduce the retarded time $u$ by letting
\begin{equation}\label{eq:Tortoise}
	t = u + r_\ast(r)\,,\qquad
	r_\ast(r) = r + 2M \log\left(
	\frac{r}{2M}-1
	\right)
\end{equation}
and in this way the metric \eqref{eq:Schwarzschild} becomes
\begin{equation}\label{eq:SchwRet}
	ds^2 = - f(r) du^2-2 du dr + r^2 d\theta^2 + r^2 (\sin\theta)^2 d\phi^2\,.
\end{equation}
Moreover,
\begin{equation}
	\left(\frac{\partial}{\partial t}\right)_{r,\theta,\phi} 
	=
	\left(\frac{\partial}{\partial u}\right)_{r,\theta,\phi} 
	,\qquad
	\left(\frac{\partial}{\partial r}\right)_{t,\theta,\phi} = 
	\left(\frac{\partial}{\partial r}\right)_{u,\theta,\phi} 
	-
	\frac{1}{f(r)} 	\left(\frac{\partial}{\partial u}\right)_{r,\theta,\phi} .
\end{equation}

Introducing instead the advanced time $v$ by letting
\begin{equation}\label{eq:Tortoisev}
	t = v - r_\ast(r),
\end{equation}
the metric  \eqref{eq:Schwarzschild} becomes
\begin{equation}\label{eq:SchwAdv}
	ds^2 = - f(r) dv^2+2 dv dr + r^2 d\theta^2 + r^2 (\sin\theta)^2 d\phi^2
\end{equation}
and one finds
\begin{equation}
	\left(\frac{\partial}{\partial t}\right)_{r,\theta,\phi} 
	=
	\left(\frac{\partial}{\partial v}\right)_{r,\theta,\phi} 
	,\qquad
	\left(\frac{\partial}{\partial r}\right)_{t,\theta,\phi} = 
	\left(\frac{\partial}{\partial r}\right)_{v,\theta,\phi} 
	+
	\frac{1}{f(r)} 	\left(\frac{\partial}{\partial v}\right)_{r,\theta,\phi} .
\end{equation}

One can check that the Penrose  tetrad introduced in \eqref{eq:Kinnersley} is properly normalized since, letting $\ell_a = (\ell,n,m,\bar{m})$ for $a=1,2,3,4$,
\begin{equation}
	\ell_a\cdot \ell_b = \left(
	\begin{matrix}
		0 & -1 & 0 & 0\\
		-1 & 0 & 0 & 0\\
		0 & 0 & 0 & 1\\
		0 & 0 & 1 & 0\\
	\end{matrix}
	\right).
\end{equation}
Moreover,
\begin{equation}\label{eq:vanishing1}
	R_{\mu\nu\alpha\beta} \ell^\mu m^\nu \ell^\alpha =0\,,\qquad
	R_{\mu\nu\alpha\beta} \ell^\mu m^\nu m^\beta = 0\,,
\end{equation}
and 
\begin{equation}\label{eq:vanishing2}
	R_{\mu\nu\alpha\beta} n^\mu \bar{m}^\nu n^\alpha = 0\,,
	\qquad
	R_{\mu\nu\alpha\beta} n^\mu \bar{m}^\nu \bar{m}^\beta =0\,,
\end{equation}
so that in particular $\Psi_0$ and $\Psi_4$ defined by \eqref{eq:defPsi4Psi0} vanish identically for this spacetime.
For completeness, let us provide the explicit expression for the spin-weighted spherical harmonics $Y_{s}^{\ell m}(\theta,\phi)$ used throughout the paper,
\begin{equation}\label{eq:SWSH}
	\begin{split}
		Y_{s}^{\ell m}(\theta,\phi)
		&=
		e^{i m \phi}
		\sqrt{\frac{(\ell+m)!(\ell-m)!(2\ell+1)}{4\pi(\ell+s)!(\ell-s)!}}
		\left(\sin\frac{\theta}{2}\right)^{2\ell}
		\\
		&\times
		\sum_{r=0}^{\ell-s}
		{\scriptsize\binom{\ell-s}{r} \binom{\ell+s}{r+s-m}}
		(-1)^{\ell+m-r-s}
		\left(
		\cot\frac{\theta}{2}
		\right)^{2r+s-m}\,.
	\end{split}
\end{equation}

\section{Derivation of source terms}
\label{app:derivation_source_terms}
In this appendix, we compute the source terms for all the spins we considered in the text.\\
From the definition of $Z_{\ell m, s}^H$ in \eqref{eq:Zinfty}, we see that we need, first of all, to analyze the harmonic components of the stress energy tensor, $T_{\ell m}$. In particular, from  \eqref{eq:Tellm} we can write these coefficients in terms of the source $\mathcal{T}$ defined in \eqref{eq:psisigma}. Indeed, thanks to the orthogonality relation of the spherical harmonics, we find that
\be
T_{\ell m}(\omega, r) = \int dt \, d\Omega \, e^{i \omega t} \, Y_{s}^{\ell m *} \, \mathcal{T} (t,r,\theta,\phi) 
\ee
with $d\Omega=d\theta \sin \theta d\phi$. For this reason we can write
\begin{align} \label{eq:expression_ZH_App}
Z_{\ell m,s}^H(\omega) & = \int dr \,\mathfrak{R}_{\text{up}}(\omega,r) \,   \Delta(r)^s \, T_{\ell m} (\omega,r) \notag \\
& =   \int dt \, dr \, d\Omega\, e^{{\rm i} \omega t}   \mathfrak{R}_{\rm up}(\omega,r) \,   Y_s^{\ell m*}(\theta , \phi)  \Delta(r)^s {\cal T} (t,r,\theta , \phi) 
\end{align}
and the final goal is showing that, for all the spins, the general expression of this quantity is given by \eqref{eq:ZbL}, that we rewrite here for convenience
\be
Z_{\ell m,s}^H(\omega) = \int d t\,e^{i\omega t - i m \phi(t)} \sum_{i=0}^2  b^i_{\ell m}\,\mathcal{L}_{i} [\mathfrak{R}_{\text{up}}(\omega, r)] \, .
\ee
Now we are going to list the main ingredients that constitute all the objects that appear and also the basic identities that will be extensively used in the computations. The sources $\mathcal{T}$ are defined in \cite{Teukolsky:1972my} and they are built from the following quantities, specified to Schwarzschild geometry
\beaq \label{eq:ingredients_teuk}
& D = \frac{\partial_t}{f(r)} +\partial_r ~, \quad  \Delta_T = \frac{1}{2} \left[\partial_t - f(r)\partial_r \right]   ~, \quad  \delta=\frac{1}{\sqrt{2} \, r} \left(\partial_\theta +\frac{i}{s_\theta} \partial_\phi\right) \nn\\
& \rho =  -\frac{1}{r} ~, \quad \beta=\frac{1}{2\sqrt{2}\, r} \cot\theta ~, \quad \pi=\tau=0 ~, \quad  \mu=-\frac{f(r)}{2r}~, \quad \gamma=\frac{M}{2r^2} \, .
\eeaq
 In addition we introduce the following operators in Schwarzschild coordinates \footnote{While instead $J_{\pm} = \partial_r -(1 \pm 1) \frac{1}{f(r)}\, \partial_u$ in retarded coordinates.}
 \begin{align} 
	L_{h} &= \partial_\theta - \frac{i}{\sin\theta}\,\partial_\phi + h \cot\theta\,,\qquad
	L_{h}^\dagger  = \partial_\theta + \frac{i}{\sin\theta}\,\partial_\phi + h \cot\theta \nn\\
	J_{\pm} &= \partial_r \mp \frac{1}{f(r)}\, \partial_t \label{intpart}
\end{align}
 and the shorthand notation
 \begin{equation}
	\mathcal{T}_{MN} = M_\alpha T^{\alpha \beta} N_\beta \,,\qquad {\cal J}_M=M_\alpha  J^{\alpha}
\end{equation}
 with $M$, $N$ some vectors. We will always use the integration by part identities
 \begin{subequations}
 \begin{align} 
\int d\Omega \, F(\theta,\phi)^* L_{-h} G(\theta,\phi) 
&=  -   \int d\Omega \,G(\theta,\phi) \left( L_{h+1}^\dagger F(\theta,\phi)\right)^* , \label{eq:integration_parts_L}\\
\int d\Omega \,  F(\theta,\phi)^* L^\dagger_{h} G(\theta,\phi) 
&=  - \int d\Omega \, G(\theta,\phi) \left(L_{1-h} F(\theta,\phi)\right)^* , \label{eq:integration_parts_Ld}\\
\int dr  F(r)   J_+ G(r) 
&= -  \int dr  \,G(r)\,  J_-   F(r)\,, \label{eq:integration_parts_J}\\
\int dt\,  e^{{\rm i} \omega t}  \partial_t G(t) 
&=  \int dt  \, e^{{\rm i} \omega t}   ( -{\rm i} \omega ) G(t)\,,  \label{eq:integration_parts_t}
\end{align}
 \end{subequations}
together with the following relations between the scalar product of the quadrivelocity of the probe $\frac{dx}{d\tau}$ and the vectors of the tetrad (due to the fact that we consider geodesics moving on the $\theta=\frac{\pi}{2}$ plane as discussed in Section~\ref{sec:geodesic})
 \begin{equation} \label{eq:contraction_quadrivalocity_vectors}
	\frac{dx}{d\tau}\cdot \ell = - \frac{E}{\mu f}\left(
	1-\frac{\dot{r}}{f}
	\right),\quad
	\frac{dx}{d\tau}\cdot n = - \frac{1}{2}\frac{E}{\mu}\left(
	1 + \frac{\dot{r}}{f}
	\right),\quad
	\frac{dx}{d\tau}\cdot \bar{m} = - \frac{i}{\sqrt{2}\,r}\frac{L}{\mu}\,.
\end{equation}
At this point we are going to analyze the computations for all the different spins. For the $s=-2$ case we report the most important passages, while for the remaining cases we consider only the main steps.
  
\subsection*{$s=-2$}  
  The Teukolsky source is given by $\mathcal{T} = 4 \pi G r^2 \mathcal{E}[T]$ (see the first line of \eqref{eq:psisigma}), where
\begin{align}
 &{\cal E}[T] = 2 (\Delta_T + 2 \gamma + \mu) [(\Delta_T - 2 \mu - \mu^*) \rho^{-4} \mathcal{T}_{\bar{m}\bar{m}} - (\delta^* - 2 \pi - 2 \beta^* - 2 \tau^*) \rho^{-4} \mathcal{T}_{n \bar{m}}]  \notag \\
 & + 2 (\delta^*+3 \pi - 2 \beta^* - \tau^*) [(\delta^*-2 \pi - \tau^*) \rho^{-4} \mathcal{T}_{nn} - (\Delta_T + 2 \gamma + 2\mu^* - 4 \mu) \rho^{-4} \mathcal{T}_{n \bar{m}}]\,.
 \end{align}
Using the expressions in \eqref{eq:ingredients_teuk} we find
%\begin{equation}\label{eq:OperatorE}
%	\begin{split}
%		\mathcal{T} 
%		&= 2 \pi G \left[2r^4 L_{-1} L_0 \mathcal{T}_{nn} 
%		+
%		\sqrt{2}\,r^3\,f(r)^2 J_+ \frac{r^2}{f(r)}\,L_{-1} \mathcal{T}_{n\bar{m}} \right.
%		\\
%		&\left. +\sqrt{2} \, r f(r)^2 J_+ \frac{r^4}{f(r)}\, L_{-1}  \mathcal{T}_{n\bar{m}}
%		+r f(r)^2J_+ r^4 J_+ r \mathcal{T}_{\bar{m}\bar{m}}  \right]
%	\end{split}
%\end{equation}
\begin{equation}\label{eq:OperatorE}
	\begin{split}
		\mathcal{T} 
		&= 4 \pi G r^2 \left[r^2 L_{-1} L_0 \mathcal{T}_{nn} 
		+
		\frac{1}{\sqrt{2}}\,r\,f(r)^2 J_+ \frac{r^2}{f(r)}\,L_{-1} \mathcal{T}_{n\bar{m}}  \right.
		\\
		&\left. + \frac{1}{r \sqrt{2}} \, f(r)^2 J_+ \frac{r^4}{f(r)}\, L_{-1}  \mathcal{T}_{n\bar{m}}
		+\frac{f(r)^2}{2 r} J_+ r^4 J_+ r \mathcal{T}_{\bar{m}\bar{m}}  \right].
	\end{split}
\end{equation}
Thanks to the following property
\begin{equation} \label{eq:intermediate_property}
	[r^\alpha ,J_{+}] = -\alpha \,r^{\alpha-1}\, , \qquad \text{where $[A,B]=AB-BA$} \,,
\end{equation}
we can write 
\begin{equation}
	\frac{\mathcal{T}}{\Delta(r)^2}
	= 4 \pi G \left[
	L_{-1} L_0 \left(
	\frac{\mathcal{T}_{nn}}{f(r)^2}
	\right)
	+ \sqrt{2} L_{-1} \left(
	J_+ r+2
	\right)
	\frac{\mathcal{T}_{n\bar{m}}}{f(r)}
	+
	\frac{1}{2}(J_+^2 r^2 + 2 J_+ r ) \mathcal{T}_{\bar{m}\bar{m}} \right]\,.
\end{equation}
When we are inside the integral, we can move the action of the $L_h$ operators to the 
%(complex conjugate of the) 
spherical harmonics thanks to the integration by parts \eqref{eq:integration_parts_L}. We can do the same with $J_+$ according to \eqref{eq:integration_parts_J} and \eqref{eq:integration_parts_t}. Since $L^\dagger_h$ and $J_-$ act on eigenfunctions of $\partial_\phi$ and $\partial_t$, their expressions reduce to
\begin{equation} \label{eq:final_expression_L_J}
	L_h^\dagger = \partial_\theta - \frac{m}{\sin\theta}+ h \cot \theta\,,
	\qquad
	J_{-} = \partial_r - \frac{i\omega}{f(r)}\,.
\end{equation}
In this way we find
\begin{equation}
	\begin{split}
		&Z_{\ell m,-2}^H (\omega) 
		= 4\pi G\int_{2M}^\infty dr \int_{-\infty}^{+\infty} dt\, d \Omega \, e^{i\omega t - i m \phi} \Big[
		\frac{1}{f(r)^2} \left(
		L_1^\dagger L_2^\dagger S_{-2}^{\ell m} \right)  \mathcal{T}_{nn} 
		\\
		&- \frac{\sqrt{2}}{f(r)}
		\left(L_2^\dagger S_{-2}^{\ell m} \right) \mathcal{T}_{n\bar m}
		(2-r J_-)
		+
		\frac{1}{2} S_{-2}^{\ell m} \,T_{\bar m \bar m}
		(r^2 J_-^2-2r J_-)
		\Big]
		\mathfrak{R_{\rm up}}(\omega ,r)\,.
	\end{split}
\end{equation}
where we have used the fact that $Y_{s}^{\ell m} = e^{i m \phi } S_{s}^{\ell m}(\theta)$ (where actually $S_{s}^{\ell m}$ are real functions). The identities
\begin{equation}\label{eq:Lids}
	L_1^\dagger L_2^\dagger S_{-2}^{\ell m}
	=
	S_{0}^{\ell m}
	\sqrt{(\ell^2-1)\ell(\ell+2)}
	\,,
	\qquad
	L_2^\dagger S_{-2}^{\ell m} 
	=
	-
	S_{-1}^{\ell m}
	\sqrt{(\ell-1)(\ell+2)}
\end{equation}
allow one to explicitly evaluate the corresponding terms.
Specializing to the stress tensor for the point particle \eqref{eq:stresspoint},
\begin{equation}
	T^{\alpha\beta} = \mu \int \frac{\delta^{(4)}(x-x(\tau))}{r^2\sin\theta}\,\frac{dx^\alpha(\tau)}{d\tau}\frac{dx^\beta(\tau)}{d\tau}\, d \tau \ ,
\end{equation} 
we have (we are on the equatorial plane, so $\theta(\tau) = \pi/2$)
\begin{align}
	\nonumber
	Z_{\ell m,-2}^H (\omega) 
	&= 4\pi G \mu \int \frac{d\tau}{r(\tau)^2} \,e^{i\omega t(\tau)-i m \phi(\tau)} \Big[
	\frac{1}{f(r)^2}
	\left(L_1^\dagger L_2^\dagger S_{-2}^{\ell m}\right)
	(\tfrac{dx}{d\tau}\cdot n)^2 
	\\
	&- \frac{\sqrt{2}}{f(r)}
	\left(L_2^\dagger S_{-2}^{\ell m}\right)
	(\tfrac{dx}{d\tau}\cdot n)(\tfrac{dx}{d\tau} \cdot \bar m)
	(2-r J_-)
	\label{eq:Zintermediate1}
	\\
	&+
	\frac{1}{2} S_{-2}^{\ell m}\,
	(\tfrac{dx}{d\tau} \cdot \bar m)^2
	(r^2 J_-^2-2r J_-)
	\Big]
	\mathfrak{R}_{\rm up}(\omega ,r(\tau))\,.
	\nonumber
\end{align}
We then change the integration variable from $\tau$ to $t$, taking into account that $d \tau = \frac{f(r) \mu}{E} dt$. Substituting the contractions of the quadrivelocity with the vectors of the tetrad via \eqref{eq:contraction_quadrivalocity_vectors}, we arrrive at
\eqref{eq:ZbL} with $b^0_{\ell m},b^1_{\ell m},b^2_{\ell m}$ and $\mathcal{L}_0,\mathcal{L}_1,\mathcal{L}_2$ given by \eqref{eq:b_sm2} and \eqref{eq:L_sm2}.

\subsection*{ $s=2$ }
The Teukolsky source is given by $\mathcal{T} = 4 \pi G r^2 \tilde{\mathcal{E}}[T]$ (see the second line of \eqref{eq:psisigma}), where
\begin{align}
 \tilde{\cal E}[T] & =  2 (\delta - 2 \beta - 4 \tau) [(\delta - \pi^*) \mathcal{T}_{\ell \ell} - (\D_T  - 2 \rho^*) \mathcal{T}_{\ell m}]  \notag \\
 & +  2 (D -4 \rho - \rho^*) [(D -\rho^*) \mathcal{T}_{mm} - (\delta - 2 \beta + 2\pi^*) \mathcal{T}_{\ell m}]\,.
 \end{align}
Using the expressions in \eqref{eq:ingredients_teuk} we find
\begin{equation}
	\begin{split}
		\mathcal{T}
		&= 4 \pi G r^2 
		\Big[
		L_{-1}^\dagger L_{0}^\dagger\, \frac{\mathcal{T}_{\ell\ell}}{r^2}
		-\frac{\sqrt{2}}{r}\left(
		J_-+\frac{2}{r}
		\right)L_{-1}^\dagger \mathcal{T}_{\ell m} 
		\\
		&-
		\left(
		J_-+\frac{5}{r}
		\right)
		\frac{\sqrt{2}}{r}
		\,L_{-1}^\dagger \mathcal{T}_{\ell m}
		+
		2 \left(
		J_-+\frac{5}{r}
		\right)
		\left(
		J_-+\frac{1}{r}
		\right)
		\mathcal{T}_{mm}
		\Big].
	\end{split}
\end{equation}
At this point, we use the analogous property of \eqref{eq:intermediate_property}, but with $J_-$ in place of $J_+$, then the integration by parts identity of $L^\dagger$ operators according to \eqref{eq:integration_parts_Ld} and the same identities for $J_-$ according to \eqref{eq:integration_parts_J} and \eqref{eq:integration_parts_t}. In this way we get that
\begin{equation} \label{eq:final_expression_L_J_2}
	L_s = \partial_\theta + \frac{m}{\sin\theta}+ s \cot \theta\,,
	\qquad
	J_{+} = \partial_r + \frac{i\omega}{f(r)}\,,
\end{equation}
and then we can use the new relations
\begin{equation}
	L_1 L_2 S_{2}^{\ell m}
	=
	S_{0}^{\ell m}
	\sqrt{(\ell^2-1)\ell(\ell+2)}
	\,,
	\qquad
	L_2 S_{2}^{\ell m} 
	=
	S_{1}^{\ell m}
	\sqrt{(\ell-1)(\ell+2)}\,.
\end{equation}
Invoking the stress tensor for the point particle, we arrive at
\begin{align}\label{eq:Zintermediate2}
	&Z_{\ell m,2}^H (\omega) 
	= 4\pi G \mu \int d\tau \,e^{i\omega t(\tau)-i m \phi(\tau)} \Big[
	r^2 f^2 (L_1 L_2 S_{2}^{\ell m})
	(\tfrac{dx}{d\tau}\cdot \ell)^2 +
	\\
	&- 4 \sqrt{2} r^2 \, f
	\left(L_2 S_{2}^{\ell m}\right)
	(\tfrac{dx}{d\tau} \cdot \ell)(\tfrac{dx}{d\tau} \cdot m)
	\left(1 + \frac{r f}{2} J_+\right)
	+ 2 r^2
	S_{2}^{\ell m}
	(\tfrac{dx}{d\tau} \cdot m)^2
	\left(f^2 r^2 J_+^2 + \right. \notag \\
	& \left. + 2 f (3 r - 2M) J_+ + 4 \left(1-\frac{2 M^2}{r^2}\right)\right) 
	\Big]
	\mathfrak{R}_{\rm up}(\omega, r(\tau))\,.
	\nonumber
\end{align}
We then change the integration variable from $\tau$ to $t$ as above and substitute the contractions of the quadrivelocity with the vectors of the tetrad via \eqref{eq:contraction_quadrivalocity_vectors}. In this way we arrrive at
\eqref{eq:ZbL} with $b^0_{\ell m},b^1_{\ell m},b^2_{\ell m}$ and $\mathcal{L}_0,\mathcal{L}_1,\mathcal{L}_2$ given by \eqref{eq:b_sp2} and \eqref{eq:L_sp2}.

 \subsection*{ $s=-1$}
The Teukolsky source is given by $\mathcal{T} = 4 \pi r^2  \mathcal{E}[J]$ (see the third line of \eqref{eq:psisigma}), where
\begin{equation}
 {\cal E}[J] = (\Delta_T + \mu) \rho^{-2} \mathcal{J}_{\bar{m}} - (\delta^*+ \pi - \tau^*) \rho^{-2} \mathcal{J}_n\,.
 \end{equation}
Using the expressions in \eqref{eq:ingredients_teuk} we find
\begin{equation}
	\begin{split}
		\mathcal{T}
		&= 4 \pi r^2 
		\Big[
		-\frac{f(r)}{2} \left(J_+ + \frac{1}{r}\right) r^2 \mathcal{J}_{\bar{m}} - \frac{1}{\sqrt{2} r} L_0 \, r^2 \mathcal{J}_n
		\Big].
	\end{split}
\end{equation}
In evaluating the piece $\mathcal{T}/\Delta(r)$ that appears in the integrand, we find that the operator $J_+$ already acts on all the objects from the left and for this reason we do not use the property \eqref{eq:intermediate_property}, which was instead important for the $s=-2$ case. As done above, we then use the integration by parts identity of the $L_0$ operator according to \eqref{eq:integration_parts_L} and the same identities for $J_+$ according to \eqref{eq:integration_parts_J} and \eqref{eq:integration_parts_t}. In this way we get that $L^\dagger_1$ and $J_-$ become as in \eqref{eq:final_expression_L_J}. By doing it we then have to use the following expression for the action of the $L^\dagger_1$ operator on the function $S_{-1}^{\ell m}$
\begin{equation}
	L^\dagger_1 S_{-1}^{\ell m}
	=
	-\sqrt{\ell (\ell+1)} S_{0}^{\ell m}\,.
\end{equation}
Invoking the current density for the point particle in \eqref{eq:J_expression}, we arrive at
\begin{align}
	&Z_{\ell m,-1}^H (\omega) 
	= 2\pi q_e \int d\tau \,e^{i\omega t(\tau)-i m \phi(\tau)} \Big[
	\frac{\sqrt{2}}{r f} \left(L^\dagger_1 S_{-1}^{\ell m}\right) \left(\tfrac{dx}{d\tau} \cdot n \right) \notag \\
	& + \frac{S^{\ell m}_{-1}}{r} \left(\tfrac{dx}{d\tau}  \cdot \bar{m} \right) \left(-1+r J_- \right)
	\Big]
	\mathfrak{R}_{\rm up}(\omega, r(\tau))\,.
\end{align}
We then change the integration variable from $\tau$ to $t$ as above and substitute the contractions of the quadrivelocity with the vectors of the tetrad via \eqref{eq:contraction_quadrivalocity_vectors}. In this way we arrive at
\eqref{eq:ZbL} with $b^0_{\ell m},b^1_{\ell m},b^2_{\ell m}$ and $\mathcal{L}_0,\mathcal{L}_1$ given by \eqref{eq:b_sm1} and \eqref{eq:L_sm1}.

\subsection*{ $s=1$ }
The Teukolsky source is given by $\mathcal{T} = 4 \pi  r^2 \tilde{\mathcal{E}}[J]$ (see the fourth line of \eqref{eq:psisigma}), where
\begin{equation}
\tilde{\cal E}[J] = (\delta - 2 \tau) \mathcal{J}_{\ell} - (D -2 \rho - \rho^*) \mathcal{J}_m\,.
 \end{equation}
Using the expressions in \eqref{eq:ingredients_teuk} we find
\begin{equation}
	\begin{split}
		\mathcal{T}
		&= 4 \pi r^2 
		\Big[
		\frac{1}{r \sqrt{2}} L^\dagger_0 \mathcal{J}_{\ell} - \left(J_- + \frac{3}{r}\right) \mathcal{J}_m
		\Big].
	\end{split}
\end{equation}
In evaluating the piece $\mathcal{T} \Delta(r)$ that appears in the integrand, we need to move the operator $J_{-}$ in such a way it acts on all the objects from the left and it is done by using the property \eqref{eq:intermediate_property} with $J_-$ in place of $J_+$. As done above, we then use the integration by parts identity of the $L^\dagger_0$ operator according to \eqref{eq:integration_parts_Ld} and the same identities for $J_-$ according to \eqref{eq:integration_parts_J} and \eqref{eq:integration_parts_t}. In this way we get that $L_1$ and $J_+$ become as in \eqref{eq:final_expression_L_J_2}. By doing it we then have to use the following expression for the action of the $L_1$ operator on the function $S_{1}^{\ell m}$
\begin{equation}
	L_1 S_{1}^{\ell m}
	=
	\sqrt{\ell (\ell+1)} S_{0}^{\ell m}\,.
\end{equation}
Invoking the current density for the point particle in \eqref{eq:J_expression}, we arrive at
\begin{align}
	&Z_{\ell m,1}^H (\omega) 
	= 4 \pi q_e \int d\tau \,e^{i\omega t(\tau)-i m \phi(\tau)} \Big[
	- \frac{r f}{\sqrt{2}} \left(L_1 S_{1}^{\ell m}\right) \left(\tfrac{dx}{d\tau} \cdot \ell \right) \notag \\
	& + r \, S^{\ell m}_{1} \left(\tfrac{dx}{d\tau} \cdot m \right) \left(1+r f J_+ \right)
	\Big]
	\mathfrak{R}_{\rm up}(\omega, r(\tau))\,.
\end{align}
We then change the integration variable from $\tau$ to $t$ as above and substitute the contractions of the quadrivelocity with the vectors of the tetrad via \eqref{eq:contraction_quadrivalocity_vectors}. In this way we arrrive at
\eqref{eq:ZbL} with $b^0_{\ell m},b^1_{\ell m},b^2_{\ell m}$ and $\mathcal{L}_0,\mathcal{L}_1$ given by \eqref{eq:b_sp1} and \eqref{eq:L_sp1}.

\subsection*{ $s=0$}
This is the simplest case. Indeed, from the relation \eqref{eq:expression_ZH_App} with $s=0$ and from the last line of \eqref{eq:psisigma}, we find that
\be
Z_{\ell m,0}^H = - 4 \pi \int dt \, dr \, d\Omega \, e^{i \omega t}\, \mathfrak{R}_{\rm up}(r) Y^{\ell m*}_0 \, r^2 \rho \, .
\ee
From the expression of the charge density for the point particle in \eqref{eq:rho_expression}, we get, always along the equatorial plane $\theta = \pi/2$
\be
Z_{\ell m,0}^H = -4 \pi q \int d\tau \, e^{i \omega t(\tau) - i m \phi(\tau)} \, S_0^{\ell m}
\ee
With the usual change of variable from $\tau$ to $t$, we exactly arrive at \eqref{eq:ZbL} with $b^0_{\ell m},b^1_{\ell m},b^2_{\ell m}$ and $\mathcal{L}_0$ given by \eqref{eq:b_s0} and \eqref{eq:L_s0}.

\section{Derivation of the fluxes}
\label{app:fluxes}

We start from the quadratic action for a perturbation $\Phi$ on a vacuum curved background with metric $g_{\alpha\beta}$,
\begin{equation}\label{eq:actionS}
	S = \int  \mathcal{L}\sqrt{-g}\,d^Dx\,.
\end{equation}
We assume that the Lagrangian density $\mathcal{L}$ depends on the perturbation $\Phi$, on its first (covariant) derivatives and on the background metric,
$
	\mathcal{L} = \mathcal{L}(\Phi, \nabla_\alpha \Phi, g_{\mu\nu})\
$.
We will be interested in the case of a scalar, $\Phi = \varphi$, a vector, $\Phi = A_\mu$, and a tensor, $\Phi=h_{\mu\nu}$, field and we do not include sources here, because we will focus on the equations of motion and on the associated fluxes away from them, either at infinity or close to the horizon. 

The variational principle applied to the action \eqref{eq:actionS} identifies
\begin{equation}
	\mathcal{E} = 
	-\nabla_\alpha \frac{\partial\mathcal{L}}{\partial \nabla_\alpha \Phi} + \frac{\partial \mathcal{L}}{\partial \Phi}
\end{equation}
as the equations of motion for $\Phi$ on the chosen background, so that $\mathcal{E}=0$ on-shell.
We consider a diffeomorphism-covariant Lagrangian and assume that $\xi^\alpha$ is a Killing vector of the background
\begin{equation}
	\mathcal{L}_\xi g_{\alpha\beta} = 0\,.
\end{equation} 
For simplicity, we focus on the case in which $\xi^\alpha$ is also a coordinate vector in the chosen chart, which will be the case of interest here, so that $\partial_\mu \xi^\alpha =0$. 
Then,
\begin{equation}\label{eq:conservedcurrent}
	j_\xi^\alpha = \frac{\partial\mathcal{L}}{\partial \nabla_\alpha \Phi} \,\mathcal{L}_\xi \Phi - \xi^\alpha \mathcal{L}
\end{equation}
is a conserved Noether current associated to the background Killing vector $\xi^\alpha$, i.e.
\begin{equation}\label{eq:conservation}
	\partial_\alpha (\sqrt{-g}\,j_\xi^\alpha) = 0
\end{equation}
on the solutions of the equations of motion. Since $\xi^\alpha$ is a coordinate Killing vector, $\mathcal{L}_\xi \Phi = \xi^\rho \partial_\rho \Phi$, and we can also introduce the Noether stress-energy tensor $\mathcal{T}^{\alpha \beta}$ by 
\begin{equation}\label{eq:NoetherSET}
	j^\alpha_\xi = - \mathcal{T}^{\alpha \beta} \xi_\beta\,,\qquad
	\mathcal{T}^{\alpha \beta}
	=
	-
	\frac{\partial\mathcal{L}}{\partial \nabla_\alpha \Phi}\,\partial^\beta\Phi + g^{\alpha\beta} \mathcal{L}\,.
\end{equation}

Considering a hypersurface $\Sigma$ with normal $n_\alpha$, we obtain the conserved charge flowing through $\Sigma$ in spacetime as follows,
\begin{equation}\label{eq:conserved_quantity}
	P_{\xi}[\Sigma] 
	= - \int j_\xi^\alpha n_\alpha \, d\Sigma
	=
	\int n_\alpha \mathcal{T}^{\alpha \beta} \xi_\beta \, d\Sigma\,,
\end{equation}
where  $d\Sigma$ is the induced measure on $\Sigma$.
Indeed, \eqref{eq:conservation} and Stokes' theorem imply that, on-shell, $P_{\xi}[\Sigma_1]=P_{\xi}[\Sigma_2]$ for any two surfaces $\Sigma_{1,2}$ with the same boundary.

We are interested in the case in which $g_{\mu\nu}$ is the Schwarzschild metric  and in the flux of conserved charges associated to energy and angular momentum at infinity and at the horizon.
Away from the black hole, we consider retarded coordinates $(u,r,\theta,\phi)$.
The total energy and angular momentum flowing through a surface at fixed $r$ are given by taking \eqref{eq:conserved_quantity} with $\xi = - \partial_u$ or $\xi = \partial_\phi$. So, the total radiated quantities at future null infinity are
\begin{equation}\label{eq:EJinftybasic}
	E_\text{rad} = \lim_{r\to\infty} \int du \oint d\Omega \, r^2\, j_{\partial_u}^r\,,
	\qquad
	J_\text{rad} = - \lim_{r\to\infty} \int du \oint d\Omega \, r^2\, j_{\partial_\phi}^r
\end{equation}
with $d\Omega = \sin\theta\,d\theta\,d\phi$.
Close to the black hole, we consider instead advanced coordinates $(v,r,\theta,\phi)$. Similarly, we consider $\xi=-\partial_v$ or $\xi=\partial_\phi$ to obtain the total energy and angular momentum falling into a surface at fixed $r$. Therefore, taking into account the surface's orientation, the total absorbed quantities at the horizon are
\begin{equation}\label{eq:EJhorizonbasic}
	E_\text{abs} = - \lim_{r\to2M} \int dv \oint d\Omega \, r^2\, j_{\partial_v}^r\,,
	\qquad
	J_\text{abs} = \lim_{r\to2M} \int dv \oint d\Omega \, r^2\, j_{\partial_\phi}^r\,.
\end{equation}

\subsubsection*{Scalar}
Let us begin from the canonically normalized scalar field, for which
\begin{equation}
	\mathcal{L} = -\frac{1}{2}\,g^{\mu\nu}\,\nabla_{\mu}\varphi \, \nabla_{\nu}\varphi\,.
\end{equation}
Then the equations of motion are
\begin{equation}
	\mathcal{E} = \Box \varphi 
\end{equation}
and the conserved current takes the form
\begin{equation}\label{eq:jxiSCALAR}
	j_\xi^\alpha = - \mathcal{T}^{\alpha \beta}\,\xi_\beta\,,
	\qquad
	\mathcal{T}^{\alpha \beta}= \nabla^\alpha\varphi\,\nabla^\beta\varphi- \frac{1}{2}\,g^{\alpha\beta}\,\nabla_\lambda\varphi \nabla^\lambda\varphi\,.
\end{equation}
We now substitute \eqref{eq:jxiSCALAR} into the general expressions \eqref{eq:EJinftybasic} and \eqref{eq:EJhorizonbasic}. Note that the term proportional to $g^{\alpha\beta}$ in \eqref{eq:jxiSCALAR} does not contribute to the current $j^r_{\xi}$ for $\xi =- \partial_u, \partial_\phi$, because the latter only involves the off-diagonal components $\mathcal{T} \indices{^{r}_{\,\,u}},\mathcal{T}\indices{^{r}_{\,\,\phi}}$. Imposing the falloff 
\begin{equation}
	\varphi(u,r,\theta,\phi) \underset{r\to\infty}{\sim} \frac{1}{r}\,\varphi^{(1)}(u,\theta,\phi) + \cdots
\end{equation}
far from the black hole, leads to 
\begin{equation}
	E_\text{rad} = \lim_{r\to\infty} \int du \oint r^2 d\Omega \, (\partial_u\varphi)^2\,,
	\qquad
	J_\text{rad} = - \lim_{r\to\infty} \int du \oint r^2d\Omega \, \partial_u \varphi\,\partial_\phi \varphi\,.
\end{equation}
Similarly, imposing that $\varphi(v,r,\theta,\phi)$ is regular as $r\to2M$ leads to
\begin{equation}
	E_\text{abs} =\lim_{r\to2M} \int dv \oint r^2 d\Omega \, (\partial_v\varphi)^2\,,
	\quad
	J_\text{abs} = - \lim_{r\to2M} \int dv \oint r^2d\Omega \, \partial_v \varphi\,\partial_\phi \varphi\,.
\end{equation}

\subsubsection*{Vector}

For the vector field,
\begin{equation}
	\mathcal{L} = - \frac{1}{4}\,F_{\mu\nu}F^{\mu\nu}\,,\qquad F_{\mu\nu} = \partial_\mu A_\nu - \partial_\nu A_{\mu}\,,
\end{equation}
the equations of motion are 
\begin{equation}
	\mathcal{E}^\alpha = \nabla_{\mu} F^{\mu\alpha}\,,
\end{equation}
and the Noether stress-energy tensor reads
\begin{equation}
	\mathcal{T}^{\alpha \beta}= F^{\alpha\mu}\partial^\beta A_{\mu} - \frac{1}{4}\,g^{\alpha\beta}\,F_{\rho\sigma} F^{\rho\sigma}\,.
\end{equation}
Imposing retarded radial gauge and falloff conditions as follows as $r\to\infty$
\begin{equation}
	A_r(u,r,\theta,\phi)=0\,,\qquad A_u (u,r,\theta,\phi) \sim \mathcal{O}(r^{-1})\,,\qquad
	A_A(u,r,\theta,\phi) \sim \mathcal{O}(r^0)\,,
\end{equation}
we obtain
\begin{equation}
	E_\text{rad} = \lim_{r\to\infty} \int du \oint r^2 d\Omega \, \partial_u A_A \partial_uA^A\,,
	\quad
	J_\text{rad} = - \lim_{r\to\infty} \int du \oint r^2d\Omega \, \partial_u A_A \partial_\phi A^A
\end{equation}
where $A=1,2$ runs over the angles and these indices are raised and lowered by $g^{AB}= r^{-2}\gamma^{AB}$, $g_{AB}=r^2\gamma_{AB}$ with $\gamma_{AB}$ the metric on the round sphere.
Imposing advanced radial gauge and falloffs as follows as $r\to2M$,
\begin{equation}
	A_r(v,r,\theta,\phi)=0\,,\qquad A_v(v,r,\theta,\phi) \sim \mathcal{O}(r-2M)\,,\qquad
	A_A(v,r,\theta,\phi) \sim \mathcal{O}((r-2M)^0)\,,
\end{equation}
we find
\begin{equation}
	E_\text{abs} = \lim_{r\to2M} \int dv \oint r^2 d\Omega \, \partial_v A_A \partial_vA^A\,,
	\quad
	J_\text{abs} = - \lim_{r\to2M} \int dv \oint r^2d\Omega \, \partial_v A_A \partial_\phi A^A\,.
\end{equation}

\subsubsection*{Tensor}
For the canonically normalized tensor field,
\begin{equation}
	\mathcal L = - \frac{1}{2}\left(
	\nabla_{\alpha} h_{\mu\nu} \nabla^{\alpha} h^{\mu\nu}
	-
	2
	\nabla_{\mu} h_{\nu\alpha} \nabla^{\nu} h^{\mu\alpha}
	-
	\nabla_\nu h \nabla^\nu h
	+
	2
	\nabla_\alpha h^{\alpha\mu}\nabla_\mu h
	\right)
\end{equation}
with $h=g^{\alpha\beta} h_{\alpha\beta}$,
\begin{equation}
	\mathcal{E}_{\mu\nu} = \Box h_{\mu\nu} - (\nabla^{\lambda}\nabla_\mu h_{\nu\lambda} + \nabla^{\lambda}\nabla_\nu h_{\mu\lambda} )
	+
	\nabla_\mu \nabla_\nu h
	-
	g_{\mu\nu}
	(
	\Box h - \nabla^\lambda \nabla^\rho h_{\lambda \rho}
	)\,,
\end{equation}
and 
\begin{equation}
	\mathcal{T}^{\alpha \beta}=(\nabla^\alpha h^{\mu\nu}-2\nabla^{\mu}h^{\nu\alpha}-g^{\mu\nu}\nabla^\alpha h+ g^{\alpha\mu}\nabla^\nu h+ g^{\mu\nu}\nabla_\lambda h^{\lambda\alpha})\partial^\beta h_{\mu\nu} + g^{\alpha\beta} \mathcal{L}\,.
\end{equation} 
We impose the retarded Bondi gauge 
\begin{equation}
	h_{rA}(u,r,\theta,\phi) =0=h_{rr}(u,r,\theta,\phi)\,,\qquad \gamma^{AB}h_{AB}(u,r,\theta,\phi) =0
\end{equation}
and the standard falloff conditions as $r\to\infty$ \cite{Barnich:2010eb}
\begin{equation}
	\begin{split}
	h_{uu}(u,r,\theta,\phi) 
	&\sim \mathcal{O}(r^{-1})\,,
	\quad
         h_{ur}(u,r,\theta,\phi) \sim \mathcal{O}(r^{-2})\,,
	 \\
	h_{uA}(u,r,\theta,\phi) 
	&\sim \mathcal{O}(r^{0})\,,
	\quad
	\hspace{2.45pt} h_{AB}(u,r,\theta,\phi) \sim \mathcal{O}(r^{1})\,,
	\end{split}
\end{equation}
and obtain
\begin{equation}
	E_\text{rad} = \lim_{r\to\infty} \int du \oint r^2 d\Omega \, \partial_u h_{AB} \partial_u h^{AB}\,,
	\quad
	J_\text{rad} = - \lim_{r\to\infty} \int du \oint r^2d\Omega \, \partial_u h_{AB} \partial_\phi h^{AB}\,.
\end{equation}
Finally, we impose the advanced Bondi gauge
\begin{equation}
	h_{rA}(v,r,\theta,\phi) =0=h_{rr}(v,r,\theta,\phi)\,,\qquad \gamma^{AB}h_{AB}(v,r,\theta,\phi) =0
\end{equation}
and the falloff conditions as $r\to2M$, that is, $f=1-2M/r\to0$,
\begin{equation}
	\begin{split}
	h_{vv}(v,r,\theta,\phi) 
	&\sim \mathcal{O}(f^{1})\,,
	\quad
	\hspace{4.53pt}
	h_{vr}(v,r,\theta,\phi) \sim \mathcal{O}(f^{1})\,,
	 \\
	h_{vA}(v,r,\theta,\phi) 
	&\sim \mathcal{O}(f^{1})\,,
	\quad
	h_{AB}(v,r,\theta,\phi) \sim \mathcal{O}(f^0)\,,
	\end{split}
\end{equation}
and obtain
\begin{equation}
	E_\text{abs} = \lim_{r\to2M} \int dv \oint r^2 d\Omega \, \partial_v h_{AB} \partial_v h^{AB}\,,
	\quad
	J_\text{abs} = - \lim_{r\to 2M} \int dv \oint r^2d\Omega \, \partial_v h_{AB} \partial_\phi h^{AB} \,.
\end{equation}
This concludes the derivation of the fluxes employed in the main text starting from Section~\ref{sec:fluxes}.
Note that $h_{\mu\nu}$ must be rescaled by $(32\pi G)^{-1/2}$ to go from the canonically normalized field employed here to the dimensionless metric fluctuation used there.

\bibliographystyle{utphys}
\bibliography{hie-4}

\end{document}